 \documentclass[11pt,a4paper]{article}
 \usepackage{jheppub}
 \usepackage{amsmath,amstext,amssymb}
 \usepackage{graphicx}
 \usepackage{feynmp}
 \usepackage{epstopdf}
 \usepackage{caption}
 \usepackage{float}
\def\be{\begin{equation}}
\def\ee{\end{equation}}
\def\ba{\begin{eqnarray}}
\def\ea{\end{eqnarray}}
 \newcommand{ \eq}[1]{Eq.~(\ref{#1})}
 \newcommand{ \fig}[1]{Fig.~(\ref{#1})}
 \newcommand{ \nn}{ \nonumber}
 \newcommand{ \bas}{ \bar{ \alpha}_S}

 \newcommand{ \Lb}{ \left(}
 \newcommand{ \Rb}{ \right)}
 \newcommand{ \h}{ \frac{1}{2}}

 \usepackage{graphics}
 \usepackage{graphicx}% Include figure files
 \usepackage{dcolumn}% Align table columns on decimal point
 \usepackage{bm}% bold math
 \usepackage{epsfig}
 \usepackage{graphicx}
 \usepackage{multirow}
 \usepackage{dcolumn}% Align table columns on decimal point
 \usepackage{graphicx,epsfig}%
 \usepackage{epstopdf}% \nofiles

\usepackage[a4paper, total={5in, 8in}]{geometry}
\begin{document}
 \title{On the energy spectrum of the electroweak Pomeron}
 \author[a]{Jochen Bartels,}%, \note{Corresponding author.}}
 \author[b,c]{Eugene Levin}
 \author[c]{Marat Siddikov}
 \affiliation[a,c]{II. Institut f \"{u}r Theoretische Physik, Universit\"{a}t Hamburg, Luruper Chaussee 149, \\
D-22761 Hamburg, Germany}
 \affiliation[b]{Department of Particle Physics, School of Physics and Astronomy,
Raymond and Beverly Sackler
 Faculty of Exact Science, Tel Aviv University, Tel Aviv, 69978, Israel}
  \affiliation[c]{Departemento de F\'isica, Universidad T\'ecnica Federico Santa Mar\'ia,  Centro Cient\'ifico- \\
Tecnol\'ogico de Valpara\'iso, Avda. Espana 1680, Casilla 110-V, Valpara\'iso, Chile }
  \emailAdd{jochen.bartels@desy.de}
   \emailAdd{leving@post.tau.ac.il,eugeny.levin@usm.cl}
     \emailAdd{marat.siddikov@usm.cl}
     \date{ \today}

% \pacs{12.38.Cy, 12.38.Aw,12.40.Vv}

% \begin{abstract}
\abstract{
In this paper we study the high energy behaviour of Electroweak Standard Model
for a nonzero Weinberg angle $\theta_{W}$. We evaluate the spectrum
of the electroweak pomeron and demonstrate that the leading intercept
is given by $\alpha _{\rm e.w.}4 \ln 2$ and does not depend on the mixing angle $\theta_{W}$. Due to its very small numerical value, we conclude that the high energy behaviour of electroweak
theory cannot be discussed without including the QCD Pomeron which, at sufficiently large 
energies, will dominate.}
%  \end{abstract}
 
  \maketitle

\section{Introduction}

High energy scattering processes described by the $SU(2)\otimes U(1)$
gauge theory of electroweak interactions, in particular the elastic
scattering of weak vector bosons, have played an important role in
unravelling the gauge structure of the electroweak sector. 
From the very beginning the Higgs mechanism was conceived as a way
to avoid power law violations of unitarity at high energies. Nevertheless,
due to logarithmic loop corrections, resummation is required in order
to understand the asymptotic high energy behaviour of the theory.
In a pure electroweak weak theory these corrections stem from the reggeization
of the electroweak gauge bosons and from the formation of the electroweak pomeron~\cite{BLP}.
When the QCD sector is taken into account, due to formation of virtual
quarks, also the QCD pomeron~\cite{KLF,BL} starts contributing~\cite{Chachamis:2003vq}
and eventually overtakes  the electroweak pomeron. This implies that the unitarization problem in the electroweak sector is still awaiting its final solution.  

Due to the rich particle content of the Standard Model, the study of the
electroweak pomeron is a challenging task. As a first step, the decoupled limit
($\theta_{w}=0$) of the spectrum of the $SU(2)$ pomeron was studied in
\cite{LLS,Levin:2015noa}, and it has been found that the inclusion
of a vector meson mass does not affect the spectrum of the Pomerons. However, it does change significantly the form of the eigenfunctions.

In this paper we extend this study to the realistic case of a nonzero
mixing angle $\theta_{w}$. \emph{A priori} it is not clear how mixing
might affect the high energy behaviour, given the fact that the Higgs
mechanism, even in the leading logarithmic approximation, introduces new contributions to the interaction kernels. As a first step, we will work in the leading logarithmic approximation and neglect effects of the runnig coupling constant. Our main (numerical) result is
that the $\theta_{w}Á\neq 0$ corrections to the eigenvalue spectrum are small, indicating
that the eigenvalues of the electroweak Pomeron coincide with our
earlier results~\cite{LLS,Levin:2015noa} for an $SU(2)$ gauge
theory.  

Our paper is organized as follows. In Section~\ref{sec:EWPom} we
give a short overview of the  BFKL approach applied to electroweak
interactions. First (section~\ref{subsec:su2}) we review the
case when the Higgs field is absent, then (section~\ref{subsec:symbreak})
we analyze how the Higgs mechanism leads to a system of coupled equations for the electroweak  pomeron \cite{BLP}, and finally (section~\ref{subsec:LargeKappa} ) we demonstrate
that the Higgs mechanism does not affect the large-transverse momentum limit
of the theory. In order to study the influence of the small-momentum region
on the physically important leading intercept, we then perform, in Section~\ref{sec:PerturbativeAnalysis}, a perturbative expansion in $\sin^{2}\theta_{W}$, making use of the smallness of the Weinberg angle. After a short review of our numerical methods developed earlier ~\cite{LLS,Levin:2015noa} we compute corrections to the leading intercept of the order $\mathcal{O}(\sin^{2}\theta_{W})$ and argue that they vanish.  After that in Section~\ref{sec:GeneralAnalysis} we reduce our system of initially 17 coupled equations for the electroweak pomeron to a single equation (which is linear in the wave function, but has a complicated  dependence on the eigenvaule $\omega$) and perform a numerical analysis 
in the lattice approximation. Finally, in Section~\ref{sec:Conclusions} we draw
our conclusions.

\section{Electroweak interactions at high energies: generalities}

\label{sec:EWPom}

\subsection{The massless $SU(2)$ pomeron}

\label{subsec:su2}

The Lagrangian of the electroweak interaction can be written as a
sum of several contributions~\cite{Agashe:2014kda} 
\begin{equation}
{\cal L}_{\mbox{ \tiny EW}}\,\,=\,\,{\cal L}_{\mbox{ \tiny gauge}}\,+\,{\cal L}_{\mbox{ \tiny fermions}}\,+\,{\cal L}_{\mbox{ \tiny Higgs boson}}\,+\,{\cal L}_{\mbox{ \tiny Yukawa}},
\label{LBE}
\end{equation}
where ${\cal L}_{\mbox{ \tiny gauge}}$ has the form 
\begin{equation}
{\cal L}_{\mbox{ \tiny gauge}}\,\,=\,\,-\,\frac{1}{4}W_{a}^{\mu\nu}\, W_{\mu\nu}^{a}\,-\,\frac{1}{4}B^{\mu\nu}\, B_{\mu\nu}\label{LBEG}.
\end{equation}
and $W_{\mu\nu}^{a}$ and $B_{\mu\nu}$ are the field strength tensors
for the gauge fileds of groups $SU(2)$ and $U(1)$ respectively.
For a moment we disregard completely all the other terms in Eq.~(\ref{LBE})
since at high energy only exchange of the vector gauge fields \cite{KOLEB}
gives the contributions to the leading $\ln(1/x)$ order of perturbative
QCD (LLA)%
\footnote{In LLA we sum all Feyman diagrams considering $\bas\ll1,\bas\ln\left(1/x\right)\,\sim\,1$
and $\bas\ln\left(Q^{2}\right)\,\ll\,1$.%
}. In LLA the massless field $W_{\mu}^{a}$ reggeizes and the behaviour
of the scattering amplitude at high energies is governed by the BFKL
evolution~%
\footnote{Throughout this paper, we will be interested in the BFKL eigenvalue
problem. Rather than dealing with BFKL-Green's functions, $G_{\omega}(q,k,k')$,
we therefore define amplitudes by convoluting the Green's function
at one end with some impact factor (which we do not need to specify).
The resulting amplitudes $\phi(\omega,q,k)$ are defined to include
one of the external momentum propagators.%
}

\begin{equation}
\phi_{W}\left(k^{2},Y=\ln\left(1/x\right)\right)\,\,=\,\,\int\frac{d\omega}{2\pi i}\, e^{\omega Y}\phi_{W}\left(\omega,k^{2}\right),\label{BFKL1}
\end{equation}
where
\begin{equation}
\omega\,\phi_{W}\left(\omega,k^{2}\right)\,\,=\,\,\bar{\alpha}_{\mbox{ \tiny e.w}}\,\Bigg\{\int\frac{dk'^{2}}{|k^{2}-k'^{2}|+k\epsilon}\,\phi_{W}\left(\omega,k'^{2}\right)\,\,\,-\,\,\ln\left(k^{2}/\epsilon^{2}\right)\,\phi_{W}\left(\omega,k^{2}\right)\Bigg\},\label{BFKL}
\end{equation}
 $\bar{\alpha}_{\mbox{ \tiny e.w}}\,\,=\,\,\frac{g^{2}}{4\,\pi}\,\frac{2}{\pi}$
for the $SU(2)$ group, the second term in~(\ref{BFKL}) stems from
vector boson reggeization, and $\epsilon\to0$ provides a regularization
near the point $k^{2}=k'^{2}$ in the emission kernel. The solution
of~(\ref{BFKL}) is well known \cite{KOLEB,BFKL}. The eigenvalues
(spectrum) of the BFKL equation is continous and is parametrized by
a variable $\nu$ which is related to eigenvalue as 
\begin{equation}
\omega_{\mbox{ \tiny e.w.BFKL}}\left(\nu\right)=\bar{\alpha}_{\mbox{ \tiny e.w}}\,\Bigg(2\psi\left(1\right)\,-\,\psi\left(\frac{1}{2}-i\nu\right)\,-\,\psi\left(\frac{1}{2}+i\nu\right)\Bigg),
\label{EWBFKL}
\end{equation}
where $\psi(z)$ is a polygamma function. The spectrum~(\ref{EWBFKL})
is limited from above by $\omega\left(\nu=0\right)\,\,=\,\,4\ln2\,\bar{\alpha}_{\mbox{ \tiny e.w}}$,
which yields for the large-$s$ behavior $s^{1+\omega\left(\nu=0\right)}.$
For all other $\nu$, the spectrum is twice degenerate, with eigenfunctions
given by 
\begin{equation}
\phi_{W}^{\mbox{ \tiny BFKL}}\left(k^{2}\right)\,\,=\,\,\left(k^{2}\right)^{-\frac{1}{2}\pm i\nu}.\label{BFKLEF}
\end{equation}
The set~(\ref{BFKLEF}) forms an orthonormalized and a complete set
of functions. The abelian fields $B_{\nu}$ do not interact with each
other in LLA and they lead to the amplitude which is proportional
to $s$~\cite{FGL}.

\subsection{Symmetry breaking and the electroweak pomeron}

\label{subsec:symbreak}

As peculiar features of the electroweak theory, due to a Higgs
mechanism gauge bosons get masses (which, naively speaking, we expect not to affect
high energy behaviour of a theory) and physical fields
$Z$ and $\gamma$ arew defined as mixtures of gauge fields $W^{3}$ and $B$.
Taking into account that only $W^{3}$ reggeizes, it is not clear
how the switch from ($W^{3},\, B$) to physical ($Z,\,\gamma$) will
affect a high energy behaviour of the theory. For this reason it is
not obvious: on the one hand, we expect that a low-energy spontaneous
symmetry breaking cannot affect the high energy amplitudes~%
\footnote{We thank Lev Lipatov for discussing this point with us.%
}. On the other hand, the Higgs field couples to both gauge fields
and generates new vertices absent in unbroken theory. The systematic
investigation of high energy scattering in the electroweak sector
in the leading log approximation was described in \cite{BLP}. It
was found that the charged gauge bosons reggeizes, and the corresponding
trajectory function is given by 
\begin{equation}
\omega_{c}(t)=\alpha_{c}(t)-1=(t-M_{W}^{2})\left(c_{W}^{2}\beta_{WZ}(t)+s_{W}^{2}\beta_{W\gamma}(t)\right)\label{eq:omega_c_0},
\end{equation}
where $c_{W}\equiv\cos\theta_{W}$, $s_{W}\equiv\sin\theta_{W}$,
$t=-q^{2}$, $q$ is a transverse momentum of a boson, and 
\begin{equation}
\beta_{ij}(t)=%g^{2}\int\frac{d^{2}k}{(2\pi)^{3}}\frac{1}{k^{2}+M_{i}^{2}}\frac{1}{(q-k)^{2}+M_{j}^{2}}=
\frac{\bar{\alpha}_{ew}}{4\pi}\int d^{2}k\frac{1}{k^{2}+M_{i}^{2}}\frac{1}{(k-q)^{2}+M_{j}^{2}}.\label{eq:beta_WW}
\end{equation}
In a neutral channel situation is more complicated since reggeization
exists for a linear combination of $Z$-boson and photon $\gamma$
which corresponds to a component $W^{3}$ of $SU(2)$ gauge field.
A general consideration in this case is complicated, and it was demonstrated
in~\cite{BLP} that it can be significantly simplified if we consider
this reggeizeable part as a separate contribution of two auxiliary
fictitious fields $n$ and 3, assuming that $Z$ and $\gamma$ do
not reggeize at all. The corresponding propagators of real and fictitious
particles are given by 
\begin{equation}
Z\,\to~~\frac{c_{W}^{2}}{\omega}\frac{1}{q^{2}+M_{Z}^{2}},\hspace{0.5cm}\gamma\,\to~~\frac{s_{W}^{2}}{\omega}\frac{1}{q^{2}},\hspace{0.5cm}n\,\to~~\frac{1}{\omega-\omega_{n}(q^{2})}\frac{1}{q^{2}+M_{W}^{2}}\hspace{0.5cm}3\,\to~~-\frac{1}{\omega}\frac{1}{q^{2}+M_{W}^{2}}\,,\label{PROP}
\end{equation}
and the neutral trajectory is defined as 
\begin{equation}
\omega_{n}(t)=\alpha_{n}(t)-1=(t-M_{W}^{2})\beta_{WW}(t),
\label{eq:omega_n_0}
\end{equation}
where $\beta_{WW}$ is given by~(\ref{eq:beta_WW}). The trajectory
function $\alpha_{n}(t)$ passes though unity at $t=M_{W}^{2}$, i.e.
neither the $Z$-boson nor the photon lie on this trajectory. In the
limit $\theta_{W}\to0$ the sum of the four exchanges (\ref{PROP})
reduces to 
\begin{equation}
\frac{1}{\omega-\omega_{n}(t)}\,\frac{1}{q^{2}+M_{W}^{2}},
\end{equation}
and $\omega_{c}$ and $\omega_{n}$ coincide. For large momenta only
the contribution of the field $n$ survives, 
\begin{equation}
\frac{c_{W}^{2}}{q^{2}+M_{Z}^{2}}+\frac{s_{W}^{2}}{q^{2}}-\frac{1}{q^{2}+M_{W}^{2}}={\cal O}\left(\frac{M_{W}^{4}}{(q^{2})^{3}}\right).\label{PROPas}
\end{equation}
When formulating coupled integral equations for the different exchange
channels, it was found to be convenient to treat the four terms in
(\ref{PROP}) as independent states: $Z,\gamma,n,3$. In the following
we will use the bracketed notations $\{Z\, n\}\,=\, Z\, n\,+n\, Z$
and$\{n\,3\}\,=\,3\, n\,+\, n\,3$. 

The corresponding eigenfunctions of BFKL evolution should satisfy
a system of coupled equations~%
\footnote{For simplicity in this paper we restrict ourselves to the forward
region $q=0$.%
} 
\begin{align}
\left(\omega-\omega_{i}(k)-\omega_{j}(k)\right)\Phi_{ij}\left(k\right) & =\int\frac{d^{2}k'}{(2\pi)^{3}}\,\sum_{i'j'\not=\gamma}K_{ij,j'j'}\frac{(-1)^{N_{3}(i',j')}c_{W}^{2N_{Z}(i',j')}\Phi_{i'j'}\left(k'\right)}{D\left(k',M_{i}\right)D\left(k',M_{j}\right)}\label{eq:B1}\\
 & +\sqrt{2}\int\frac{d^{2}k'}{(2\pi)^{3}}'\, K_{ij,cc}\left(k,k'\right)\frac{\Phi_{cc}\left(k'\right)}{D\left(k',M_{W}\right)^{2}}\label{eq:K_ij_ij}
\end{align}
\begin{align}
\left(\omega-2\omega_{c}(k)\right)\Phi_{cc}\left(k\right) & =\sqrt{2}\int\frac{d^{2}k'}{(2\pi)^{3}}\,\sum_{i'j'}K_{cc,ij}\frac{(-1)^{N_{3}(i,j)}c_{W}^{2N_{Z}(i,j)}s_{W}^{2N_{\gamma}(i,j)}\Phi_{ij}\left(k'\right)}{D\left(k',M_{i}\right)D\left(k',M_{j}\right)}\label{eq:B2}\\
 & +\int\frac{d^{2}k'}{(2\pi)^{3}}\, K_{cc,cc}\left(k,k'\right)\frac{\Phi_{cc}\left(k'\right)}{D\left(k',M_{W}\right)^{2}},
\label{eq:K_cc_cc}
\end{align}
where indices $i,\, j,\, i',\, j'$, unless stated otherwise, run
over above-mentioned neutral states $Z,\,\gamma,\, n,\,3$; and $N_{m}(i,j)$
stands for the number of times ``$m$'' appears among its arguments
(so for example $N_{Z}(Z,Z)=2$, $N_{Z}(Z,\gamma)=1$, ...). The factors
$(-1)^{N_{3}(i,j)}c_{W}^{2N_{Z}(i,j)}s_{W}^{2N_{\gamma}(i,j)}$appear
from numerators of propagators, and in denominators which stem from
propagators~(\ref{PROP}) we use shorthand notations $D\left(k,\, M_{i}\right)\equiv k^{2}+M_{i}^{2}$.
For the sake of brevity we will use an abbreviation $D(k)\equiv D\left(k,\, M_{W}\right)$,
and $M_{W}\equiv M$. The corresponding kernels have a form 
\begin{eqnarray}
K_{ij,j'j'} & = & \frac{g^{2}M_{W}^{2}}{2c_{W}^{2N_{Z}(i,j,i',j')}}\theta\left(i,j,i',j'\not=\gamma\right),\label{kijij}\\
K_{ij,cc}\left(k,\, k'\right) & = & g^{2}\left(-M_{ij}^{2}+\frac{\left(D\left(k,\, M_{i}\right)+D\left(k,\, M_{j}\right)\right)D\left(k'\right)}{D(k-k')}\right)\label{kijcc}\\
 & = & g^{2}\left(-M_{ij}^{2}+2K_{em}\left(k,k'\right)+\frac{\left(M_{i}^{2}+M_{j}^{2}-2M_{W}^{2}\right)D\left(k'\right)}{D(k-k')}\right),\nonumber \\
K_{em}\left(k,k'\right) & = & g^{2}\frac{D\left(k\right)D\left(k'\right)}{D(k-k')}=g^{2}K_{em}\left(k,k',M_{W}\right),\label{eq:K_em_def}\\
K_{em}\left(k,k',M\right) & = & g^{2}\frac{D\left(k\right)D\left(k'\right)}{D(k-k',M)},\\
M_{ij}^{2} & = & M_{i}^{2}+M_{j}^{2}-\frac{M_{i}^{2}M_{j}^{2}}{2M_{W}^{2}},\\
K_{cc,cc} & = & g^{2}\left(-M_{W}^{2}+D\left(k\right)D\left(k'\right)\left(\frac{c_{W}^{2}}{D\left(k-k',\, M_{Z}\right)}+\frac{s_{W}^{2}}{D\left(k-k',\,0\right)}\right)\right)\\
 & \equiv & g^{2}\left(-M_{W}^{2}+\left(c_{W}^{2}K_{em}\left(k,k',M_{Z}\right)+s_{W}^{2}K_{em}\left(k,k',M_{Z}\right)\right)\right),\nonumber 
\end{eqnarray}
where $\theta$-function in~(\ref{kijij}) reflects the fact that
the kernel $K_{ij,i'j'}$ vanishes if any of the indices is $\gamma$~%
\footnote{This happens because in the leading order over $\alpha_{EW}$ there
are no terms with a direct coupling of a Higgs boson to massless photon.%
}. In these notations the trajectories~(\ref{eq:omega_c_0},\ref{eq:omega_n_0})
can be rewritten as 
\begin{eqnarray}
\omega_{i}(k) & = & -\delta_{i,n}g^{2}\int\frac{d^{2}k'}{(2\pi)^{3}}\frac{D(k)}{D(k')D(k-k')},\label{eq:omega_i}\\
\omega_{c}(k) & = & -g^{2}\int\frac{d^{2}k'}{(2\pi)^{3}}\frac{D(k)}{D(k')}\left(\frac{c_{W}^{2}}{D(k-k',\, M_{Z})}+\frac{s_{W}^{2}}{D(k-k',\,0)}\right).\label{eq:omega_c}
\end{eqnarray}

\subsection{Large momentum asymptotics}

\label{subsec:LargeKappa}

At large transverse momenta $k\gg M$, we expect that the theory should
not depend on a value of mixing angle $\theta_{W}$. We will show
explicitly that in this region the system of equations reduces to
equation for $SU(2)$ pomeron, supporting the intuitive expectation
that the low-energy symmetry breaking does not affect the high energy
asymptotic behaviour of the scattering amplitude. In the region of
large transverse momenta $k,\, k',|k-k'|\gg M$ we can neglect all
the masses, so the emission kernels take a form

\begin{eqnarray}
K_{ij,cc}\left(k,\, k'\right) & \approx & 2g^{2}K^{\mbox{ \tiny BFKL}}\left(k,\, k'\right)\label{kijcc-1}\\
K_{cc,cc} & \approx & g^{2}K^{\mbox{ \tiny BFKL}}\left(k,\, k'\right)
\end{eqnarray},
where the BFKL kernel is given by 
\begin{equation}
K^{\mbox{ \tiny BFKL}}\left(k,\, k'\right)\,\,=\,\, g^{2}\frac{k^{2}{k'}^{2}}{(k-k')^{2}}.\label{KERBFKL}
\end{equation}
The equation~(\ref{eq:B1}) simplifies to
\begin{align}
\left(\omega-\omega_{i}(k)-\omega_{j}(k)\right)\Phi_{ij}\left(k\right) & \approx2\sqrt{2}g^{2}\int\frac{d^{2}k'}{(2\pi)^{3}}'\,\frac{K^{\mbox{ \tiny BFKL}}\left(k,\, k'\right)}{D\left(k',M_{W}\right)^{2}}\Phi_{cc}\left(k'\right)\label{BFKLLIM}
\end{align}
and implies that neutral fields $\Phi_{ij}$ differ only due to Regge
trajectories in the lhs. of~(\ref{BFKLLIM}). In particular, using~(\ref{eq:omega_i},\ref{eq:omega_c}),
we may get that at large momenta 
\begin{align}
\Phi_{ZZ} & \approx\Phi_{\gamma\gamma}\approx\Phi_{33}\approx\Phi_{Z\gamma}\approx\Phi_{Z3}\approx\Phi_{\gamma3},\label{eq:large_ZZ}\\
\Phi_{Zn} & \approx\Phi_{3n}\approx\Phi_{Z3}.\label{eq:large_Zn}
\end{align}

Combining this with~(\ref{PROPas}), we can see that in~(\ref{PROPas})
the contribution of all components with indices $Z,\,\gamma$ or $3$
mutually cancel (decouple) at large $k$, \emph{i.e.} in (\ref{eq:B2})
only a field ``$n$'' contributes. Also, we can notice that there
is no terms which depend explicitly on $\theta_{W}$. If we assume
that $\sqrt{2}\Phi_{nn}(k)=\Phi_{cc}$ and does not depend on azimulthal
angle, after redefinition 
\begin{equation}
\phi_{W}^{\mbox{\tiny BFKL}}\left(k\right)\,=\,\,\frac{\Phi_{cc}\left(k\right)}{D\left(k,M\right)}\label{eq:norm}
\end{equation}
we may reproduce~(\ref{BFKL}). It is known that the large-$k$ asymptotics
of solutions of~(\ref{BFKL}) is the same as in massive case~(\ref{BFKLEF}),
\begin{equation}
\phi_{ij}^{\mbox{\tiny BFKL}}\left(k\right)\,\propto\,\left(k^{2}\right)^{-\frac{1}{2}\pm i\nu},\label{eq:phi_asymp}
\end{equation}
such that
\begin{equation}
\Phi_{ij}\left(k\right)\,\propto\,\left(k^{2}\right)^{\frac{1}{2}\pm i\nu}.\label{eq:Phi_asymp}
\end{equation}

The fact that $\Phi_{ij}\left(k\right)$ grows at large $k$ justifies
omission of contact terms in~(\ref{eq:B1}) and finalizes our proof
that for asymptotically large $k$ the eigenfunctions and eigenvalues
coincide with $SU(2)$ pomeron, independently of the value of $\theta_W$.

A crucial assumption which was implicitly used in this section is
that if we consider large momentum $k$, then in the integrals in~(\ref{eq:B1},\ref{eq:B2})
the dominant contribution also comes from large $k'$ and large $|k-k'|$.
Potentially the latter assumption might be violated near $k\approx k'$
where the emission kernel~(), giving rise to a strong sensitivity
to $\theta_{W}$. For this reason in the following Section~\ref{sec:PerturbativeAnalysis}
we perform an analysis of $\mathcal{O}\left(\theta_{w}\right)$ corrections
and demonstrate explicitly that it does not affect the spectrum. Additionally,
the proof given above could be invalidated by localized solutions
which fall off rapidly at large $k$ as $\phi_{i,j}\propto1/k^{2}$
or faster (see Appendix~\ref{sub:discrete} for a particular example
of such solution). If such solutions exist and have intercept $\omega\,>\,\omega^{\mbox{ \tiny BFKL}}\left(\nu=0\right)$,
potentially they could significantly change the evolution of the BFKL
spectrum. For this reason in Section~\ref{sec:GeneralAnalysis} we
perform a general analysis in the lattice and demonstrate that there
is no such solutions.

\section{Perturbative analysis in the small-$\theta_{W}$ limit}

\label{sec:PerturbativeAnalysis} In this section we develop a systematic
perturbative expansion of~(\ref{eq:B1},\ref{eq:B2}) over the parameter
$s_{w}^{2}=\sin^{2}\theta_{W}$ up to the first order of perturbation
theory. In addition to explicit dependence on Weinberg angle in~(\ref{PROP}).
we should also take into account the $\theta_{w}$ dependence in $Z$-boson
mass, 
\begin{equation}
M_{Z}^{2}=\frac{M_{W}^{2}}{c_{W}^{2}}\approx M_{W}^{2}\left(1+s_{w}^{2}\right).
\end{equation}

The trajectory $\omega_{c}\left(q^{2}\right)$ in this limit may be
expanded as 
\begin{equation}
\omega_{c}\left(q^{2}\right)\approx\omega_{n}\left(q^{2}\right)+\, s_{W}^{2}\,\Delta\omega_{c}\left(q^{2}\right),
\end{equation}
where
\begin{equation}
\Delta\omega_{c}\left(q^{2}\right)=\bar{\alpha}_{ew}\left(q^{2}-M^{2}\right)\int\frac{d^{2}k}{4\pi}\frac{M^{4}}{k^{2}D(k)^{2}D(q-k)}.
\label{DOM}
\end{equation}
From now on we use a notation $\Delta$ for the corrections proportional
to $s_{w}^{2}$. The kernel $K_{ij,i'j'}$ given in~(\ref{kijij})
may be expanded as
\begin{equation}
K_{ij,i'j'}\approx\frac{g^{2}M_{W}^{2}}{2}\theta\left(i,j,i',j'\not=\gamma\right)\left(1+N_{Z}(i,j,i',j')s_{W}^{2}\right).
\end{equation}

In a similar fashion, using the expansion of $M_{ij}^{2}$
\begin{eqnarray}
M_{ZZ}^{2} & = & \frac{3}{2}M^{2}+s_{w}^{2}M^{2}\nonumber \\
M_{Z\gamma}^{2} & = & M^{2}+s_{w}^{2}M^{2}\nonumber \\
M_{Zn}^{2} & = & \frac{3}{2}M^{2}+\frac{1}{2}s_{w}^{2}M^{2}.
\end{eqnarray}
\begin{eqnarray}
M_{\gamma\gamma}^{2} & = & 0\nonumber \\
M_{nn}^{2} & = & \frac{3}{2}M^{2}\nonumber \\
M_{\gamma n}^{2} & = & M^{2}\nonumber \\
M_{n\gamma}^{2} & = & M^{2}
\end{eqnarray},
we may obtain for the kernels with account of $\mathcal{O}\left(s_{w}^{2}\right)$
corrections 
\begin{eqnarray}
K_{ZZ;cc} & = & g^{2}\left((-q^{2}-\frac{3}{2}M^{2})+\frac{D(k)D(q-k')+D(q-k)D(k')}{D(k-k')}\right)\nonumber \\
 &  & +s_{w}^{2}g^{2}\left(-M^{2}+M^{2}\frac{D(k)+D(q-k)}{D(k-k')}\right),\label{eq:K_ZZcc}
\end{eqnarray}
\begin{eqnarray}
K_{Z\gamma;cc} & = & g^{2}\left(-q^{2}-M^{2}+\frac{D(k)D(q-k')+(q-k)^{2}D(k')}{D(k-k')}\right)\nonumber \\
 &  & +s_{w}^{2}g^{2}\left(-M^{2}+M^{2}\frac{D(q-k')}{D(k-k')}\right)\label{eq:K_Zgcc}
\end{eqnarray}
\begin{eqnarray}
K_{\gamma Z;cc} & = & g^{2}\left(-q^{2}-M^{2}+\frac{k^{2}D(q-k')+D(q-k)D(k')}{D(k-k')}\right)\nonumber \\
 &  & +s_{w}^{2}g^{2}\left(-M^{2}+M^{2}\frac{D(k')}{D(k-k')}\right)\label{eq:K_gZcc}
\end{eqnarray}
\begin{eqnarray}
K_{Zn;cc} & = & g^{2}\left(-q^{2}-\frac{3}{2}M^{2}+\frac{D(k)D(q-k')+D(q-k)D(k')}{D(k-k')}\right)\nonumber \\
 &  & +s_{w}^{2}g^{2}\left(-\frac{M^{2}}{2}+M^{2}\frac{D(q-k')}{D(k-k')}\right)\label{eq:K_Zncc}
\end{eqnarray}
\begin{eqnarray}
K_{nZ;cc} & = & g^{2}\left(-q^{2}-\frac{3}{2}M^{2}+\frac{D(k)D(q-k')+D(q-k)D(k')}{D(k-k')}\right)\nonumber \\
 &  & +s_{w}^{2}g^{2}\left(-\frac{M^{2}}{2}+M^{2}\frac{D(k')}{D(k-k')}\right)\label{eq:K_nZcc}
\end{eqnarray}
 
\begin{equation}
K_{\gamma\gamma;cc}=g^{2}\left(-q^{2}+\frac{k^{2}D(q-k')+(q-k)^{2}D(k')}{D(k-k')}\right)\label{GGCC}
\end{equation}
\begin{equation}
K_{\gamma n;cc}=g^{2}\left(-q^{2}-M^{2}+\frac{k^{2}D(q-k')+D(q-k)D(k')}{D(k-k')}\right)
\end{equation}
\begin{equation}
K_{n\gamma;cc}=g^{2}\left(-q^{2}-M^{2}+\frac{D(k)D(q-k')+(q-k)^{2}D(k')}{D(k-k')}\right)\label{NGCC}
\end{equation}
\begin{equation}
K_{nn;cc}=g^{2}\left(-q^{2}-\frac{3}{2}M^{2}+\frac{D(k)D(q-k')+D(q-k)D(k')}{D(k-k')}\right)\label{K-nn;cc}
\end{equation}
 \begin{equation}
K_{cc;ij}(q,k,k')=K_{ij;cc}(q,k',k).\label{CCNN}
\end{equation}
\begin{eqnarray}
K_{cc;cc} & = & g^{2}\left(-q^{2}-M^{2}+\frac{D(k)D(q-k')+D(q-k)D(k')}{D(k-k')}\right)\nonumber \\
 &  & +s_{w}^{2}g^{2}M^{4}\frac{D(k)D(q-k')+D(q-k)D(k')}{(k-k')^{2}D(k-k')^{2}}.\label{eq:K_cccc}
\end{eqnarray}

In what follows it is convenient to introduce a shorthand notation
$\Delta{\cal K}$ for all $\mathcal{O}\left(s_{W}^{2}\right)$-corrections,
both due to kernels and propagators. In the case of a $\Delta{\cal K}_{cc;cc}$
component, we should also add an $\mathcal{O}\left(s_{W}^{2}\right)$-contribution
from the Regge trajectory
\begin{equation}
\Delta{\cal K}_{cc;cc}=\left(\Delta K_{cc;cc}+\left(\Delta\omega(-k^{2})+\Delta\omega(-(q-k)^{2})\right)\delta^{(2)}(k-k')\right)\frac{1}{D(k')D(q-k')}.\label{DCCCC}
\end{equation},
and it is straightforward to verify that there is a cancellation of
the $s$-channel photon pole. The other components of operator $\Delta{\cal K}$
are given explicitly as
\begin{equation}
\Delta{\cal K}_{\gamma\gamma,cc}=0,\quad\Delta{\cal K}_{nn,cc}=0,\quad\Delta{\cal K}_{\gamma n,cc}=0,
\end{equation}
\begin{equation}
\Delta{\cal K}_{cc;\gamma n}=\Delta{\cal K}_{cc;\gamma Z},\quad\Delta{\cal K}_{cc;n\gamma}=\Delta{\cal K}_{cc;Z\gamma}
\end{equation}

\begin{eqnarray}
\Delta{\cal K}_{cc;ZZ} & = & g^{2}\left(2(q^{2}+M^{2})+(q^{2}+\frac{3}{2}M^{2})\Big[\frac{M^{2}}{D(k')}+\frac{M^{2}}{D(q-k')}\Big]\right.\nonumber \\
 &  & \left.-\frac{M^{2}}{D(k-k')}\Big[\frac{D(k)}{D(k')}D(q-k')+\frac{D(q-k)}{D(q-k')}D(k')\Big]\right.\nonumber \\
 &  & \left.-2\frac{D(k)D(q-k')+D(q-k)D(k')}{D(k-k')}\right)\frac{1}{D(k')D(q-k')}
\end{eqnarray}
\begin{eqnarray}
\Delta K_{cc;Zn} & = & g^{2}\left(q^{2}+M^{2}+(q^{2}+\frac{3}{2}M^{2})\frac{M^{2}}{D(k')}\right.\nonumber \\
 &  & \left.+\frac{M^{2}}{D(k-k')}\Big[D(k)-\frac{D(k)}{D(k')}D(q-k')\Big]\right.\nonumber \\
 &  & \left.-\frac{D(k)D(q-k')+D(q-k)D(k')}{D(k-k')}\right)\frac{1}{D(k')D(q-k')}
\end{eqnarray}
\begin{eqnarray}
\Delta{\cal K}_{cc;nZ} & = & g^{2}\left(q^{2}+M^{2}+(q^{2}+\frac{3}{2}M^{2})\frac{M^{2}}{D(q-k')}\right.\nonumber \\
 &  & \left.+\frac{M^{2}}{D(k-k')}\Big[D(q-k)-\frac{D(q-k)}{D(q-k')}D(k')\Big]\right.\nonumber \\
 &  & \left.-\frac{D(k)D(q-k')+D(q-k)D(k')}{D(k-k')}\right)\frac{1}{D(k')D(q-k')}
\end{eqnarray}
\begin{equation}
\Delta{\cal K}_{cc;\gamma Z}=g^{2}\left(-q^{2}-M^{2}+\frac{D(k)D(q-k')+D(q-k){k'}^{2}}{D(k-k')}\right)\frac{1}{{k'}^{2}D(q-k')}\label{CCGZ}
\end{equation}
\begin{equation}
\Delta{\cal K}_{cc;Z\gamma}=g^{2}\left(-q^{2}-M^{2}+\frac{D(k)(q-k')^{2}+D(q-k)D(k')}{D(k-k')}\right)\frac{1}{D(k')(q-k')^{2}}.
\label{CCZG}
\end{equation}

Finally the transitions neutral$\to$neutral. Here we have corrections
both from the kernels and from the propagators of the $Z$-boson and
of the photon. There are no corrections from the trajectory functions,
so they only contribute to the leading order diagonal kernels $nn\to nn$.
In the following we list the corrections for the different channels:
\begin{eqnarray}
\Delta{\cal K}_{ZZ;ZZ} & = & g^{2}\frac{M^{2}}{2}\left(2-\frac{M^{2}}{D(k')}-\frac{M^{2}}{D(q-k')}\right)\frac{1}{D(k')D(q-k')}\label{d-ZZ;ZZ}\\
\Delta{\cal K}_{nZ;ZZ} & = & g^{2}\frac{M^{2}}{2}\left(1-\frac{M^{2}}{D(k')}-\frac{M^{2}}{D(q-k')}\right)\frac{1}{D(k')D(q-k')}\label{d_nz;ZZ}\\
\Delta{\cal {\cal K}}_{nn;ZZ} & = & g^{2}\frac{M^{2}}{2}\left(-\frac{M^{2}}{D(k')}-\frac{M^{2}}{D(q-k')}\right)\frac{1}{D(k')D(q-k')}\label{d_nn;ZZ}
\end{eqnarray}
\begin{eqnarray}
\Delta{\cal K}_{ZZ;Zn} & = & g^{2}\frac{M^{2}}{2}\left(2-\frac{M^{2}}{D(k')}\right)\frac{1}{D(k')D(q-k')}\label{d-ZZ;Zn}\\
\Delta{\cal {\cal K}}_{nZ;Zn} & = & g^{2}\frac{M^{2}}{2}\left(1-\frac{M^{2}}{D(k')}\right)\frac{1}{D(k')D(q-k')}\label{d-nZ;Zn}\\
\Delta{\cal K}_{nn;Zn} & = & g^{2}\frac{M^{2}}{2}\left(-\frac{M^{2}}{D(k')}\right)\frac{1}{D(k')D(q-k')}\label{d-nn;Zn}
\end{eqnarray}
\begin{eqnarray}
\Delta{\cal K}_{ZZ;nn} & = & 2g^{2}\frac{M^{2}}{2}\label{DZZNN}\\
\Delta{\cal {\cal K}}_{Zn;nn} & = & g^{2}\frac{M^{2}}{2}\label{eq:DZNNN}
\end{eqnarray}.
The remaining kernels are easily obtained using a symmetry $\Delta K_{ij,i'j'}=\Delta K_{ji,i'j'}=\Delta K_{ij,j'i'}$.

\subsection{The case of zero Weinberg angle, $\theta_{W}=0$}

\label{subsec:Generalities}

In this section we study the eigenvalues for the case $\theta_{W}=0$.
In this limit the set of equations~(\ref{eq:B1},\ref{eq:B2}) simplifies
considerably, since the photon field decouples from the other equations,
and the other neutral fields have the same mass. Besides, all the
kernels $K$ no longer distinguish these states, and the only differences
among the neutral states is due to the nonzero Regge trajectory of
$n$. In this limit the system~~(\ref{eq:B1},\ref{eq:B2}) reduces
to a system of seven coupled equations
\begin{eqnarray}
\omega\Phi_{ZZ}\left(k\right) & = & \frac{g^{2}}{2}\int\frac{d^{2}k'}{(2\pi)^{3}}\frac{1}{D^{2}\left(k'\right)}\left(\sum_{ij\not=\gamma}(-1)^{N_{3}}\Phi_{ij}\left(k'\right)-3\sqrt{2}\Phi_{cc}\left(k'\right)\right)\nonumber \\
 & + & 2\sqrt{2}\int\frac{d^{2}k'}{(2\pi)^{3}}\frac{1}{D^{2}\left(k'\right)}\, K_{em}\left(k,k'\right)\Phi_{cc}\left(k'\right),\label{eq:ZZ}\\
\left(\omega-\omega_{n}(k)\right)\Phi_{Zn}\left(k\right) & = & \frac{g^{2}}{2}\int\frac{d^{2}k'}{(2\pi)^{3}}\frac{1}{D^{2}\left(k'\right)}\left(\sum_{ij\not=\gamma}(-1)^{N_{3}}\Phi_{ij}\left(k'\right)-3\sqrt{2}\Phi_{cc}\left(k'\right)\right)\nonumber \\
 & + & 2\sqrt{2}\int\frac{d^{2}k'}{(2\pi)^{3}}\frac{1}{D^{2}\left(k'\right)}\, K_{em}\left(k,k'\right)\Phi_{cc}\left(k'\right),\\
\omega\Phi_{Z3}\left(k\right) & = & \frac{g^{2}}{2}\int\frac{d^{2}k'}{(2\pi)^{3}}\frac{1}{D^{2}\left(k'\right)}\left(\sum_{ij\not=\gamma}(-1)^{N_{3}}\Phi_{ij}\left(k'\right)-3\sqrt{2}\Phi_{cc}\left(k'\right)\right)\nonumber \\
 & + & 2\sqrt{2}\int\frac{d^{2}k'}{(2\pi)^{3}}\frac{1}{D^{2}\left(k'\right)}\, K_{em}\left(k,k'\right)\Phi_{cc}\left(k'\right),\\
\omega\Phi_{33}\left(k\right) & = & \frac{g^{2}}{2}\int\frac{d^{2}k'}{(2\pi)^{3}}\frac{1}{D^{2}\left(k'\right)}\left(\sum_{ij\not=\gamma}(-1)^{N_{3}}\Phi_{ij}\left(k'\right)-3\sqrt{2}\Phi_{cc}\left(k'\right)\right)\nonumber \\
 & + & 2\sqrt{2}\int\frac{d^{2}k'}{(2\pi)^{3}}\frac{1}{D^{2}\left(k'\right)}\, K_{em}\left(k,k'\right)\Phi_{cc}\left(k'\right),
\end{eqnarray}
\begin{eqnarray}
\left(\omega-\omega_{n}(k)\right)\Phi_{n3}\left(k\right) & = & \frac{g^{2}}{2}\int\frac{d^{2}k'}{(2\pi)^{3}}\frac{1}{D^{2}\left(k'\right)}\left(\sum_{ij\not=\gamma}(-1)^{N_{3}}\Phi_{ij}\left(k'\right)-3\sqrt{2}\Phi_{cc}\left(k'\right)\right)\nonumber \\
 & + & 2\sqrt{2}\int\frac{d^{2}k'}{(2\pi)^{3}}\frac{1}{D^{2}\left(k'\right)}\, K_{em}\left(k,k'\right)\Phi_{cc}\left(k'\right)\\
\left(\omega-2\omega_{n}(k)\right)\Phi_{nn}\left(k\right) & = & \frac{g^{2}}{2}\int\frac{d^{2}k'}{(2\pi)^{3}}\frac{1}{D^{2}\left(k'\right)}\left(\sum_{ij\not=\gamma}(-1)^{N_{3}}\Phi_{ij}\left(k'\right)-3\sqrt{2}\Phi_{cc}\left(k'\right)\right)\nonumber \\
 & + & 2\sqrt{2}\int\frac{d^{2}k'}{(2\pi)^{3}}\frac{1}{D^{2}\left(k'\right)}\, K_{em}\left(k,k'\right)\Phi_{cc}\left(k'\right)\\
\end{eqnarray}
\begin{eqnarray}
\left(\omega-2\omega_{n}(k)\right)\Phi_{cc}\left(k\right) & = & -g^{2}\int\frac{d^{2}k'}{(2\pi)^{3}}\frac{1}{D^{2}\left(k'\right)}\left(\frac{3\sqrt{2}}{2}\sum_{ij\not=\gamma}(-1)^{N_{3}}\Phi_{ij}\left(k'\right)+\Phi_{cc}\left(k'\right)\right)+\nonumber \\
 & + & 2\sqrt{2}\int\frac{d^{2}k'}{(2\pi)^{3}}\frac{1}{D^{2}\left(k'\right)}K_{em}\left(k,k'\right)\sum_{n_{i}n_{j}\not=\gamma}(-1)^{N_{3}}\Phi_{ij}\left(k'\right)\nonumber \\
 & + & 2\int\frac{d^{2}k'}{(2\pi)^{3}}\frac{1}{D^{2}\left(k'\right)}K_{em}\left(k,k'\right)\Phi_{cc}\left(k'\right).
\label{eq:cc}
\end{eqnarray}
There are three more decoupled equations containing a photon: 
\begin{eqnarray}
\left(\omega-\omega_{n}(k)\right)\Phi_{n\gamma}\left(k\right) & = & -g^{2}\sqrt{2}\int\frac{d^{2}k'}{(2\pi)^{3}}\frac{1}{D^{2}\left(k'\right)}\Phi_{cc}\left(k'\right)+2\sqrt{2}\int\frac{d^{2}k'}{(2\pi)^{3}}\frac{1}{D^{2}\left(k'\right)}\, K_{em}\left(k,k'\right)\Phi_{cc}\left(k'\right),\nonumber \\
\omega\Phi_{Z\gamma}\left(k\right) & = & -g^{2}\sqrt{2}\int\frac{d^{2}k'}{(2\pi)^{3}}\frac{1}{D^{2}\left(k'\right)}\Phi_{cc}\left(k'\right)+2\sqrt{2}\int\frac{d^{2}k'}{(2\pi)^{3}}\frac{1}{D^{2}\left(k'\right)}\, K_{em}\left(k,k'\right)\Phi_{cc}\left(k'\right),\nonumber \\
\omega\Phi_{3\gamma}\left(k\right) & = & -g^{2}\sqrt{2}\int\frac{d^{2}k'}{(2\pi)^{3}}\frac{1}{D^{2}\left(k'\right)}\Phi_{cc}\left(k'\right)+2\sqrt{2}\int\frac{d^{2}k'}{(2\pi)^{3}}\frac{1}{D^{2}\left(k'\right)}\, K_{em}\left(k,k'\right)\Phi_{cc}\left(k'\right),\nonumber \\
\\
\omega\Phi_{\gamma\gamma}\left(k\right) & = & 2\sqrt{2}\int\frac{d^{2}k'}{(2\pi)^{3}}\frac{1}{D^{2}\left(k'\right)}K_{em}\left(k,k'\right)\Phi_{cc}\left(k'\right).
\end{eqnarray}
 As we can see from~(\ref{eq:ZZ}- \ref{eq:cc}), there are several
coinciding components of $\Phi_{ij}$ which we denote as~%
\footnote{In contrast to a result found in~\cite{BLP}, we can see that not
all neutral fields in~(\ref{eq:Phi_1_def},\ref{eq:Phi_2_def}) coincide
due to nonzero Regge trajectory $\omega_{n}(k)$. In the limit $\theta_{W}=0$
this does not affect the eigenvalues, but for the $\mathcal{O}\left(s_{W}^{2}\right)$
corrections this difference is important.%
} 
\begin{eqnarray}
\Phi_{ZZ}\left(k\right) & = & \Phi_{Z3}\left(k\right)=\Phi_{33}\left(k\right)=\Phi_{1}\left(k\right)
\label{eq:Phi_1_def}\\
\Phi_{Zn}\left(k\right) & = & \Phi_{n3}\left(k\right)=\frac{\omega}{\omega-\omega_{n}(k)}\Phi_{1}\left(k\right).
\label{eq:Phi_2_def}
\end{eqnarray}

Using~(\ref{eq:Phi_1_def},\ref{eq:Phi_2_def}), the sum over neutral
fields in~(\ref{eq:ZZ}- \ref{eq:cc}) reduces to 
\begin{equation}
\sum_{ij\not=\gamma}(-1)^{N_{3}}\Phi_{ij}\left(k'\right)=\Phi_{nn}\left(k'\right),
\end{equation}
and the full system~(\ref{eq:Phi_1_def},\ref{eq:Phi_2_def}) reduces
to a simple system of just two coupled equations. With the substitutions (after the angular integration )
\be
g^{2}\frac{d^{2}k'}{(2\pi)^{3}} \to g^{2}\frac{d\kappa}{8\pi^{2}} \to \frac{g^{2}}{4\pi}\,\frac{d\kappa}{2\pi}=\frac{\bar{\alpha}_{\mbox{ \tiny e.w}}}{4}d\kappa
\label{angular-int}
\ee
we find:
\begin{eqnarray}
\left(\omega-2\omega\left(k\right)\right)\Phi_{nn}\left(k\right) & = & \frac{g^{2}}{2}\int\frac{d^{2}k'}{(2\pi)^{3}}\frac{1}{D^{2}\left(k'\right)}\left(\Phi_{nn}\left(k'\right)-3\sqrt{2}\Phi_{cc}\left(k'\right)\right)\nonumber \\
 & + & 2\sqrt{2}\int\frac{d^{2}k'}{(2\pi)^{3}}\frac{K_{em}\left(k,\, k'\right)\Phi_{cc}\left(k'\right)}{D^{2}\left(k'\right)},\label{eq:Phi_nn}\\
\left(\omega-2\omega\left(k\right)\right)\Phi_{cc}\left(k\right) & = & -g^{2}\int\frac{d^{2}k'}{(2\pi)^{3}}\frac{1}{D^{2}\left(k'\right)}\left(\frac{3\sqrt{2}}{2}\Phi_{nn}\left(\kappa'\right)+\Phi_{cc}\left(\kappa'\right)\right)\label{eq:Phi_cc}\\
 &  & +2\sqrt{2}g^{2}\int\frac{d^{2}k'}{(2\pi)^{3}}\frac{K_{em}\left(k,\, k'\right)\Phi_{nn}\left(k'\right)}{D^{2}\left(k'\right)}+2g^{2}\int\frac{d^{2}k'}{(2\pi)^{3}}\frac{K_{em}\left(k,\, k'\right)\Phi_{cc}\left(k'\right)}{D^{2}\left(k'\right)}.\nonumber 
\end{eqnarray}
As it was discussed in~\cite{BLP}, this system of equations~(\ref{eq:Phi_nn},
\ref{eq:Phi_cc}) has two solutions, with isospin zero and two. Since
we are mostly interested in the leading intercept, in what follows,
we will consider only the isospin zero case, for which a solution has a form
\begin{equation}
\left(\begin{array}{c}
\Phi_{nn}\left(k\right)\\
\Phi_{cc}\left(k\right)
\end{array}\right)=\left(\begin{array}{c}
1\\
\sqrt{2}
\end{array}\right)\Phi\left(k\right).
\end{equation}
We thus end up with a $SU(2)$ BFKL equation for $\Phi\left(k\right)$ 
\begin{equation}
\left(\omega-2\omega(k)\right)\Phi\left(k\right)=-g^{2}\frac{5M_{W}^{2}}{2}\int\frac{d^{2}k'}{(2\pi)^{3}}\,\frac{\Phi\left(k'\right)}{D^{2}\left(k'\right)}+4\int\frac{d^{2}k'}{(2\pi)^{3}}\,\frac{K_{em}\left(k,k'\right)\Phi\left(k'\right)}{D^{2}\left(k'\right)}.
\label{eq:Phi_BFKL}
\end{equation}
which up to a renormalization of the field $\Phi$ coincides with~(\ref{BFKL})
which has been analyzed in detail in~\cite{LLS}. The field $\Phi_{1}$
defined in~(\ref{eq:Phi_1_def}) is related to $\Phi$ as 
\begin{equation}
\Phi_{1}\left(k\right)=\frac{\omega-2\omega\left(k\right)}{\omega}\Phi\left(k\right)=\left(1-\frac{2\omega\left(k\right)}{\omega}\right)\Phi\left(k\right),\label{eq:Phi_1_sol}
\end{equation}
and for the states with photons we may cast our result into the form 
\begin{eqnarray}
\Phi_{n\gamma}\left(k\right) & = & \frac{\omega-2\omega(k)}{\omega-\omega(k)}\Phi\left(k\right)+\frac{g^{2}}{2\left(\omega-\omega(k)\right)}\int\frac{d^{2}k'}{(2\pi)^{3}}\frac{\Phi\left(k'\right)}{D^{2}\left(k'\right)}\\
\Phi_{Z\gamma}\left(k\right)=\Phi_{3\gamma}\left(k\right) & = & \frac{\omega-2\omega(\kappa)}{\omega}\Phi\left(k\right)+\frac{g^{2}}{2\omega}\int\frac{d^{2}k'}{(2\pi)^{3}}\frac{\Phi\left(k'\right)}{D^{2}\left(k'\right)}\\
\Phi_{\gamma\gamma}\left(k\right) & = & \frac{\omega-2\omega(\kappa)}{\omega}\Phi\left(k\right)+\frac{5g^{2}}{2\omega}\int\frac{d^{2}k'}{(2\pi)^{3}}\frac{\Phi\left(k'\right)}{D^{2}\left(k'\right)}.
\end{eqnarray}

\subsection{Lattice solutions of the equations for $\theta_{W}=0$ }

\subsubsection{The method}

\label{subsec:NumMethod} \label{sec:EValues}In this section we analyze
the solutions in the lattice using the approach of~\cite{Levin:2015noa}.
Since the solution $\Phi$ of~(\ref{eq:Phi_BFKL}) at large momenta
is growing as~(\ref{eq:Phi_asymp}), it is more convenient to change
a normalization as in~(\ref{eq:norm}) and will work with a field
$\phi$ which is decreasing at large momenta. Since we are interested
only in solutions which do not depend explicitly on azimuthal angle,
it is convenient to perform the angular integrations and to introduce
the dimensionless variables $\kappa\,\,=\,\, k^{2}/M_{W}^{2}$ and
$\kappa'\,=\, k'^{2}/M_{W}^{2}$. Such transition in the integrals
in~(\ref{eq:Phi_BFKL}) reduces to the substitution (\ref{angular-int})
The kernel $K_{em}\left(k,k'\right)$ after angular averaging becomes
to 
\begin{equation}
K_{0}\left(\kappa,\kappa'\right)=g^{2}\frac{\left(\kappa+1\right)\left(\kappa'+1\right)}{\sqrt{(\kappa-\kappa')^{2}+2(\kappa+\kappa')+1}},
\end{equation}
and Eq. (\ref{eq:Phi_BFKL}) simplifies to 
\begin{equation}
\omega\,\phi\left(\kappa\right)\,\,=\,\,\int d\kappa'K\left(\kappa,\kappa'\right)\,\phi\left(\kappa'\right),
\label{NS1}
\end{equation}
where 
\begin{eqnarray}
 &  & K\left(\kappa,\kappa'\right)\,\,=\bar{\alpha}_{\mbox{ \tiny e.w}}\left(\frac{1}{\sqrt{(\kappa-\kappa')^{2}+2(\kappa+\kappa')+1}}-\underbrace{\frac{5}{2}\frac{1}{\kappa+1}\frac{1}{\kappa'+1}}_{{\rm contact\, term}}\right.\label{KKAPPA}\\
 &  & -\left.\frac{\kappa+1}{\sqrt{\kappa}\sqrt{\kappa+4}}\ln\frac{\sqrt{\kappa+4}+\sqrt{\kappa}}{\sqrt{\kappa+4}-\sqrt{\kappa}}\,\delta\left(\kappa\,-\,\kappa'\right)\right).\nonumber 
\end{eqnarray}
For our numerical studies we use a logarithmic grid in both variables
$\kappa$ and $\kappa'$ with $N+1$ nodes, 
\begin{eqnarray}
\kappa_{n} & = & \kappa_{min}\exp\left(\frac{n}{N}\,\ln\left(\kappa_{max}/\kappa_{min}\right)\right),\quad n=0,...,N,\label{NS2}
\end{eqnarray}
where the values of $\kappa_{min},\,\kappa_{max}$ were set to $\kappa_{min}=10^{-40},\kappa_{max}=10^{80}$,
and $N=1024$. In this grid (\ref{NS1}) takes a form of linear matrix
eigenvalue problem:
\begin{align}
\omega\phi_{n}\,\, & =\,\,\sum_{m=0}^{N}{\cal K}_{nm}\,\,\phi_{m},\label{NS3}\\
\phi_{n} & \equiv\phi\left(\kappa_{n}\right),\\
{\cal K}_{nm} & \equiv K\left(\kappa_{n},\kappa{}_{m}\right)\kappa{}_{m}\,\left(\frac{1}{N}\,\left(\kappa_{max}/\kappa_{min}\right)\right).
\end{align}
In Ref. \cite{LLS} we solved (\ref{NS3}) both for the massive~(\ref{eq:Phi_BFKL})
and for the massless BFKL equation. For the latter case the eigenvalue spectrum and the eigenfunctions obtained from the lattice approximation coincide,  with a very good precision,
with the well-known analytical results~(\ref{BFKLEF},\ref{EWBFKL}), provided
we replace the continuous parameter $\nu$ by the discrete lattice parameter 
\begin{equation}
\nu_{n}=\frac{2.9\, n}{\ln\left(\kappa_{\mbox{max}}/\kappa_{\mbox{min}}\right)}.
\end{equation}
We view this as a test of our lattice approximation.  

\begin{figure}
\centering \includegraphics[scale=0.9]{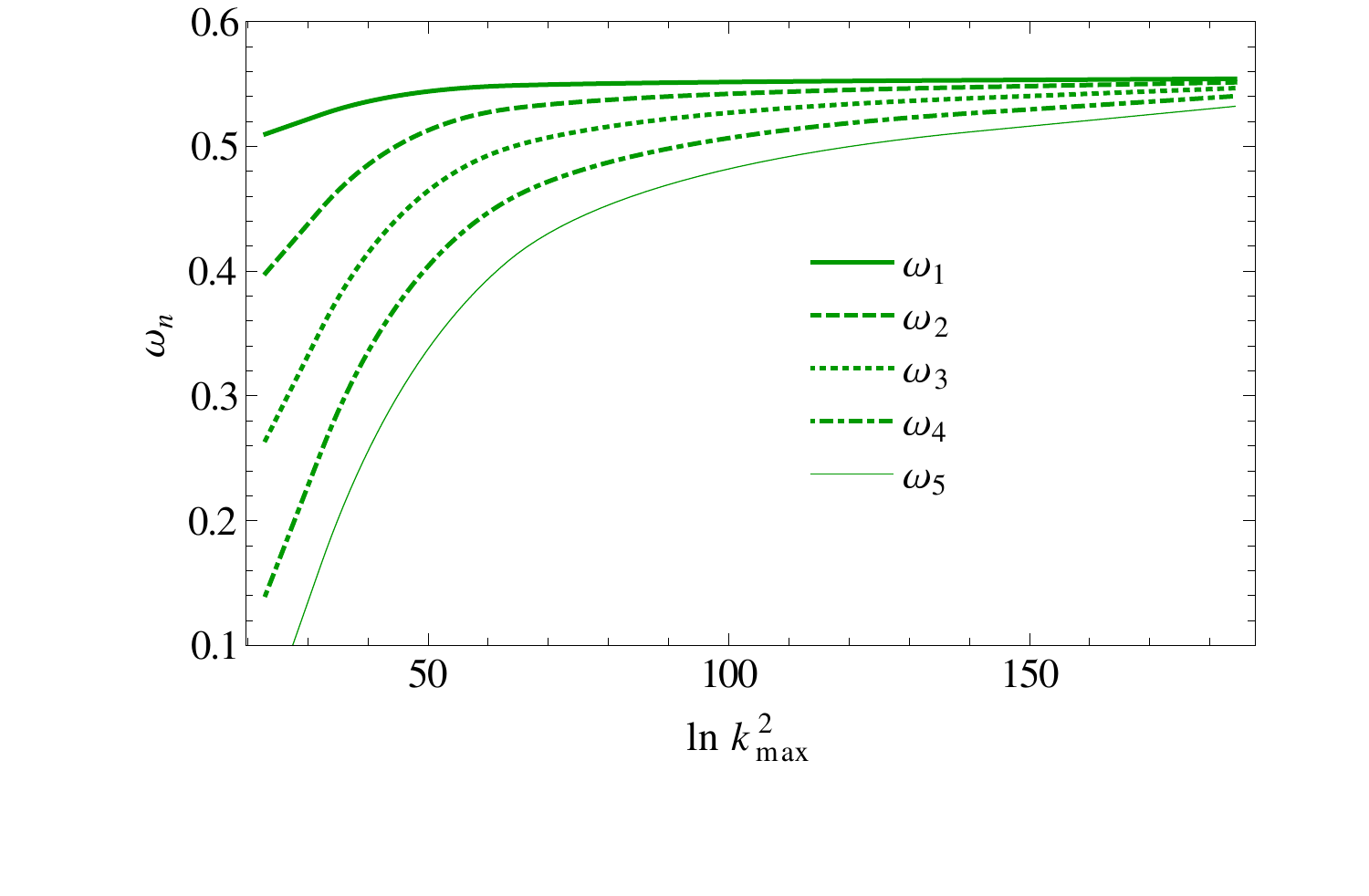} \protect\caption{\label{fig:rootsDep}Dependence of eigenvalues on the ultraviolet cutoff
$\kappa_{max}$. Only the first five eigenvalues are shown; eigenvalues with larger
$n$ uniformly fill in the band $\omega\lesssim\omega_{0}$.}
\end{figure}

For the massive case, the positive eigenvalues are described very well by
the same expression with $\nu_n$ being replaced by  
\begin{equation}
\nu_{n}^{(M)}=\frac{2.9\, n}{\ln\left(\kappa_{\mbox{max}}/\left(\kappa_{\mbox{min}}\,+\, M^{2}\right)\right)}.
\end{equation}
The first twenty eigenvalues are given in the Table~\ref{tab:dircetBFKL}.
The lattice value of the leading intercept differs from its analytical
result $\omega_{BFKL}/\bar{\alpha}_{\mbox{ \tiny e.w}}\,\,=\,\,4\ln2\,\,\,\approx\,\,2.77$
by $3\times10^{-5}$, which illustrates a very high precision of the
chosen method.
\begin{table}[h]
\begin{centering}
\begin{tabular}{|c|c|c|c|c|c|c|c|}
\hline 
Root \#  & $\omega_{n}/\bar{\alpha}_{\mbox{ \tiny e.w}}$  & Root \#  & $\omega_{n}/\bar{\alpha}_{\mbox{ \tiny e.w}}$  & Root \#  & $\omega_{n}/\bar{\alpha}_{\mbox{ \tiny e.w}}$  & Root \#  & $\omega_{n}/\bar{\alpha}_{\mbox{ \tiny e.w}}$ \tabularnewline
\hline 
1  & 2.77  & 6  & 2.61  & 11  & 2.27  & 16  & 1.825 \tabularnewline
\hline 
2  & 2.755  & 7  & 2.555  & 12  & 2.185  & 17  & 1.73 \tabularnewline
\hline 
3  & 2.735  & 8  & 2.49  & 13  & 2.1  & 18  & 1.635 \tabularnewline
\hline 
4  & 2.7  & 9  & 2.425  & 14  & 2.01  & 19  & 1.54 \tabularnewline
\hline 
5  & 2.66  & 10  & 2.35  & 15  & 1.915  & 20  & 1.445 \tabularnewline
\hline 
\end{tabular}
\par\end{centering}
\protect\protect\caption{\label{tab:dircetBFKL}The first twenty roots of the massive BFKL equation
(\ref{eq:Phi_BFKL}).}
\end{table}
\\
The eigenfunctions with positive intercepts can be parametrized as
\begin{eqnarray}
\phi_{n}^{\mbox{ \tiny(approx)}}\left(\kappa\right)\,\, & = & \,\,\frac{\alpha\left(n\right)}{\sqrt{\kappa+4}}\sin\Big(\nu_{n}^{(M)}\, Ln\left(\kappa\right)+\varphi_{n}\Big),\label{APPWF}\\
Ln\left(\kappa\right) & \equiv & \,\ln\Big(\frac{\sqrt{\kappa\,+\,4}\,\,+\,\,\sqrt{\kappa}}{\sqrt{\kappa\,+\,4}\,\,-\,\,\sqrt{\kappa}}\Big),\nonumber 
\end{eqnarray}
where $\varphi_{n}\,=b_{\phi}\,\nu_{n}^{(M)}$ with $b_{\phi}\approx1.865$.
In the continuum limit we should replace the discrete variable $\nu_{n}^{(M)}$
by the continuous variable $\nu.$ As we can see from the Figure~(\ref{fig:rootsDep}),
indeed the transition to the continuum spectrum (limit $\kappa_{max}\to\infty$)
is smooth.

\subsection{Corrections proportional to $\sin^2 \theta_{W}$}
\label{sub:sw2corr}

In this section we will estimate the corrections
of the order $\propto s_{W}^{2}$ to the leading intercept. A small
value of the parameter $\left(s_{W}^{2}\right)_{{\rm phys}}\,~\equiv\,\sin^{2}\theta_{W}=0.234\pm0.0013$
implies that a perturbative expansion should converge rapidly. While
application of the perturbation theory for the shift of band in a
continuum spectrum is not well-defined, in the lattice the spectrum
is discrete, so an ordinary perturbation theory is applicable provided
first order corrections are smaller than the distance between the
neighbor states. In this section we assume this, and show that the
corresponding correction to the intercept explicitly vanishes. In
the first order of perturbation theory, the shift of the leading intercept
is given by 
\begin{equation}
\Delta\omega_{1}\,\,=\,\,\frac{\int\, d\kappa\, d\kappa'\,\phi_{1}\left(\kappa\right)\,\Delta{\cal K}\left(\kappa,\kappa'\right)\,\phi_{1}\left(\kappa'\right)}{\Vert\phi_{1}\Vert^{2}},
\label{PE1}
\end{equation}
where the kernels $\Delta{\cal K}$ were introduced earlier in~(\ref{DCCCC}-\ref{eq:DZNNN}),
and the norm $\Vert\phi_{n}\Vert$ is defined as 
\begin{equation}
\Vert\phi_{1}\Vert^{2}\,=\,\int d\kappa\,\vert\phi_{1}\left(\kappa\right)\vert^{2}.
\label{NORM}
\end{equation}

Using the parametrization of $\phi_{1}$ from~(\ref{APPWF}), we
can see that the norm~(\ref{NORM}) is growing as a function of upper
cutoff as $\propto\ln\kappa_{max}$, which is a manifestation of the
fact that we work with wave functions of continuum spectrum. For this
reason we should only focus on the large-$\kappa_{{\rm max}}$ behaviour
of the different contributions to numerator. As we can see from (\ref{eq:K_ZZcc}-\ref{eq:K_Zncc}),
(\ref{eq:K_cccc}), all the contributions to $\Delta{\cal K}$ from
expansion of masses in kernels $\Delta K$ and Regge trajectories
$\Delta\omega_{c}$ have an additional suppression $\mathcal{O}\left(M^{2}/k^{2}\right)$
and thus lead to finite contributions to numerator (vanishing contributions
to $\Delta\omega_{1}$) in the limit%
\footnote{See Appendix~\ref{sub:NumericalLattice} for a more detailed explanation
and numerical estimates.%
} $\kappa_{max}\to\infty$. Similarly, all the contributions due to
expansion of mass $M_{Z}$ in propagators do not affect~$\Delta\omega_{1}$. 

More complicated is evaluation of the $\mathcal{O}\left(s_{W}^{2}\right)$
terms which stem from numerators of~(\ref{PROP}) which do not have
any additional power suppression. Approximating all kernels in (\ref{eq:K_ZZcc}-\ref{eq:K_Zncc}),
(\ref{eq:K_cccc}) with $K_{em}\left(k,k'\right)$ and using 
\begin{equation}
\int d\kappa'K_{em}\left(\kappa,\kappa'\right)\,\phi_{1}\left(\kappa'\right)\,\,\approx\,\,\left(\omega_{1}\,-\,2\omega_{n}\left(\kappa\right)\right)\,\phi_{1}\left(\kappa\right)+\mathcal{O}\left(\frac{M^{2}}{\kappa}\right),
\label{KEMOM}
\end{equation}
we obtain for the leading contributions to the numerator of~(\ref{PE1})
\begin{eqnarray}
\int d\kappa\,\phi_{1}\left(\kappa\right)\int d\kappa'K_{em}\left(\kappa,\kappa'\right)\,\phi_{1}\left(\kappa'\right)\,\, & = & \,\,\int d\kappa\phi_{1}\left(\kappa\right)\,\left(\omega_{1}\,-\,2\omega_{n}\left(\kappa\right)\right)\,\phi_{1}\left(\kappa\right)\label{PE2}\\
 & \approx & 2\int\, d\kappa\,\phi_{1}^{2}\left(\kappa\right)\,\omega\left(\kappa\right)\, d\kappa\,\approx\,2\int\, d\kappa\ln\kappa\,\phi_{n}^{2}\left(\kappa\right).\nonumber 
\end{eqnarray}
Separately each of these contributions leads to a logarithmically
divergent correction to eigenvalue 
\begin{equation}
\frac{\int\, d\kappa\, d\kappa'\,\phi_{1}\left(\kappa\right)\, K_{em}\left(\kappa,\kappa'\right)\,\phi_{1}\left(\kappa'\right)}{\Vert\phi_{1}\Vert^{2}}\,\,\propto\,\ln\kappa_{{\rm max}}\gg\,\,1.\label{PE3}
\end{equation}
However, in the sum over all components there are strong cancellations
of all such contributions. Indeed, as we can see from (\ref{eq:K_ZZcc}-\ref{eq:K_Zncc})
and (\ref{eq:K_nZcc}), such corrections might appear only from $K_{cc,ij}$~%
\footnote{$K_{cccc}$ also contains $K_{em}$, but the propagator of the charged
field does not contain $s_{w}^{2}$%
}, which in the large momentum limit has a structure 
\begin{equation}
\Delta K_{cc,ij}\left(\kappa,\kappa'\right)\approx4s_{W}^{2}\eta_{ij}K_{em}\left(\kappa,\,\kappa'\right)\left(1+\mathcal{O}\left(\frac{M^{2}}{\kappa}\right)\right),\label{eq:KPR}
\end{equation}
where
\begin{equation}
\eta_{Z\gamma}=\eta_{n\gamma}=\eta_{3\gamma}=-\eta_{ZZ}=-\eta_{Zn}=-\eta_{Z3}=1.
\end{equation}

After the summing all terms in the last line and taking into account
the large momentum asymptotic relations~(\ref{eq:large_ZZ},\ref{eq:large_Zn})
for the components of $\phi_{ij}$, we can see that there is a full
cancellation of all leading terms. The $\mathcal{O}\left(\frac{M^{2}}{\kappa}\right)$-corrections
lead to finite contributions to the numerator (vanishing contribution
to $\Delta\omega$). As a consequence, the correction~(\ref{PE1})
decreases as $1/\ln\left(\kappa_{{\rm max}}\right)$ in the infinite
lattice limit. This finishes a proof that $\mathcal{O}\left(\sin^{2}\theta_{W}\right)$
term vanishes in the $\kappa_{max}\to\infty$ limit. These cancellations
are a mere consequence of the fact that the large momentum limit of
the theory with Higgs mechanism is the same as for the pure $SU(2)$ pomeron.

\subsection{Corrections proportional to $\sin^{4}\theta_{W}$ and the continuum limit}

The corrections of the order of $\mathcal{O}\left(s_{W}^{4}\right)$
are relevant not only for academic interest but also because they
allow us to verify that the discrete spectrum perturbation theory
is applicable to analysis of discretized eigenvalues in the lattice.
There are two sources of such corrections. The corrections which stem
from $\mathcal{O}\left(s_{W}^{4}\right)$-expansion terms in the kernels
and propagators can be estimated from~(\ref{PE1}). From analysis
of the coefficients in front of $K_{em}$ in kernel components $K_{cc,ZZ}$,
$K_{\gamma Z,cc}$ and $K_{\gamma\gamma,cc}$, we may repeat the line
of reasoning of a previous section and demonstrate that large $\mathcal{O}\left(s_{W}^{4}\right)$-corrections
which stem from the numerators of propagators in $ZZ$, $\gamma Z$
and $\gamma\gamma$ states vanish.

More complicated is the structure of the $\mathcal{O}\left(s_{W}^{4}\right)$-corrections
in the second order of a perturbation theory which are given by 
\begin{equation}
\Delta^{(2)}\omega=\sum_{i,i\neq1}\frac{\left(\Delta\omega_{1i}\right)^{2}}{\omega_{1}\,-\,\omega_{i}},
\label{PE26}
\end{equation}
where we have introduced a shorthand notation for the transition matrix element
\begin{equation}
\Delta\omega_{1i}\,\,=\,\,\frac{\int\, d\kappa\, d\kappa'\,\phi_{1}\left(\kappa\right)\,\Delta K\left(\kappa,\kappa'\right)\,\phi_{i}\left(\kappa'\right)}{\Vert\phi_{i}\Vert\,\Vert\phi_{i}\Vert},
\label{PE27}
\end{equation}
and $\Vert\phi_{i}\Vert$ is defined in~(\ref{NORM}).

Following the analysis of the first order perturbation theory, we
derive that $\Delta\omega_{1i}$ behaves as $1/\ln\left(\kappa_{{\rm max}}\right)$
in the infinite lattice limit. The splitting of levels in denominator
of~(\ref{PE26}) for the leading intercept behaves as%
\footnote{Note that this estimate is valid only for the leading intercept. For
all other eigenvalues $\omega_{j}-\omega_{i}\propto1/\ln\left(\kappa_{max}/\left(\kappa_{min}+1\right)\right),$
so the ratio~(\ref{PE26})%
} $\omega_{1}-\omega_{i}\propto1/\ln^{2}\left(\kappa_{max}/\left(\kappa_{min}+1\right)\right)$
due to $\mathcal{O}\left(\nu^{2}\right)$ behaviour of~(\ref{EWBFKL})
near $\nu=0$. As a consequence, the ratio~(\ref{PE26}) is stable
in the limit $\kappa_{{\rm max}}\to\infty$ and deserves a special
attention. As shown in Appendix~\ref{sub:NumericalLattice}, numerically
the ratio is small, which justifies application of perturbative expansion
to discretized lattice spectrum. We see two sources of the numerical
smallness in the each term of (\ref{PE26}): the large value of the
second derivative $D\,\,=|\chi''(\nu)_{\nu=0}|=\,\,28\zeta(3)\approx33.6$
and the oscillatory behavior of the integrand in the numerator of~(\ref{PE27})
(which is a manifestation of the orthogonality of $\phi_{1}$ and
$\phi_{i}$ when $\Delta K\approx{\rm const}$).

\section{Lattice analysis for nonzero $\theta_{W}$}

\label{sec:GeneralAnalysis} 

\label{subsec:GenLattice} In this section we construct a general
solution of the Weinberg-Salam model with nonzero $\theta_{W}$. We
demonstrate that the system~(\ref{eq:B1},\ref{eq:B2}) can be reduced
to a single integral equation~\ref{eq:B2-2} for the function $\phi_{cc}$.
We do not make any assumptions about smallness of Weinberg angle $\theta_{W}$
nor select any restricted kinematics in momentum space. We start our analysis
from the observation that the kernel $K_{ij,j'j'}$ in (\ref{eq:B1})
may be factorized into two parts, as 
\begin{equation}
K_{ij,j'j'}=g^{2}\frac{M_{W}^{2}}{2c_{W}^{2N_{Z}(i,j,i',j')}}=g^{2}\frac{M_{W}^{2}}{2c_{W}^{2N_{Z}(i,j)}c_{W}^{2N_{Z}(i',j')}}\equiv g^{2}\frac{M_{i}^{2}M_{j}^{2}}{2M_{W}^{2}}\frac{1}{c_{W}^{2N_{Z}(i',j')}}.
\label{eq:K4}
\end{equation}
We notice that $c_{W}^{2N_{Z}(i',j')}$ in the denominator of~(\ref{eq:K4})
cancels against similar factors in the numerator of~(\ref{eq:B1}), which allows to 
cast the latter into the form 
\begin{eqnarray}
 &  & \left(\omega-N_{n}(i,j)\omega_{n}(k)\right)\Phi_{ij}\left(k\right)=\label{eq:B1-2}\\
 & = & g^{2}\frac{M_{i}^{2}M_{j}^{2}}{2M_{W}^{2}}\int\frac{d^{2}k'}{(2\pi)^{3}}\,\varphi_{\Sigma}\left(k'\right)+\sqrt{2}\int\frac{d^{2}k'}{(2\pi)^{3}}\,\frac{K_{ij,cc}\left(k,k'\right)\Phi_{cc}\left(k'\right)}{D\left(k',\, M\right)^{2}}\nonumber 
\end{eqnarray}
or 
\begin{eqnarray}
\Phi_{ij}\left(k\right) & = & \left(\omega-N_{n}(i,j)\omega_{n}(k)\right)^{-1}\label{eq:B1-3}\\
 & \times & \left[g^{2}\frac{M_{i}^{2}M_{j}^{2}}{2M_{W}^{2}}\int\frac{d^{2}k'}{(2\pi)^{3}}\,\varphi_{\Sigma}\left(k'\right)+\sqrt{2}\int\frac{d^{2}k'}{(2\pi)^{3}}\,\frac{K_{ij,cc}\left(k,k'\right)\Phi_{cc}\left(k'\right)}{D\left(k',\, M\right)^{2}}\right],\nonumber 
\end{eqnarray}
where 
\begin{eqnarray}
\varphi_{\Sigma}(k) & \equiv & \sum_{ij\not=\gamma}(-1)^{N_{3}(i,j)}\frac{\Phi_{ij}(k)}{D\left(k,\, M_{i}\right)D\left(k,\, M_{j}\right)}=\label{eq:PhiSigma}\\
 & = & \sum_{ij\not=\gamma}(-1)^{N_{3}(i,j)}\left(\left(\omega-N_{n}(i,j)\omega_{n}(k)\right)D\left(k,\, M_{i}\right)D\left(k,\, M_{j}\right)\right)^{-1}\nonumber \\
 & \times & \left[g^{2}\frac{M_{i}^{2}M_{j}^{2}}{2M_{W}^{2}}\int\frac{d^{2}k'}{(2\pi)^{3}}\,\varphi_{\Sigma}\left(k'\right)+\sqrt{2}\int\frac{d^{2}k'}{(2\pi)^{3}}\,\frac{K_{ij,cc}\left(k,k'\right)\Phi_{cc}\left(k'\right)}{D\left(k',\, M\right)^{2}}\right].\nonumber 
\end{eqnarray}
Equation(\ref{eq:PhiSigma}) contains, on the rhs, $\varphi_{\Sigma}\left(k'\right)$ only as part of the integral  $\int d^{2}k'\,\varphi_{\Sigma}\left(k'\right)$. So we take the integral over $k$ of both 
sides and solve for $N_{\Sigma}\left[\Phi_{cc}\right]\equiv\int\frac{d^{2}k}{(2\pi)^{3}}\,\varphi_{\Sigma}$:
\begin{eqnarray}
 &  & N_{\Sigma}\left[\Phi_{cc}\right]\equiv\int\frac{d^{2}k}{(2\pi)^{3}}\,\varphi_{\Sigma}(k)=\label{eq:PhiSigmaMoment}\\
 & = & \left[\sqrt{2}\int\frac{d^{2}k}{(2\pi)^{3}}\int\frac{d^{2}k'}{(2\pi)^{3}}\,\sum_{ij\not=\gamma}\frac{(-1)^{N_{3}(i,j)}K_{ij,cc}\left(k,k'\right)\Phi_{cc}\left(k'\right)}{\left(\omega-N_{n}(i,j)\omega_{n}(k)\right)D\left(k,\, M_{i}\right)D\left(k,\, M_{j}\right)D\left(k',\, M\right)^{2}}\right]\times\nonumber \\
 & \times & \left[1-g^{2}\sum_{ij\not=\gamma}\int\frac{d^{2}k}{(2\pi)^{3}}\frac{(-1)^{N_{3}(i,j)}}{\left(\omega-N_{n}(i,j)\omega_{n}(k)\right)D\left(k,\, M_{i}\right)D\left(k,\, M_{j}\right)}\frac{M_{i}^{2}M_{j}^{2}}{2M_{W}^{2}}\right]^{-1}.\nonumber 
\end{eqnarray}
For a given $\Phi_{cc}$ from (\ref{eq:PhiSigma},\ref{eq:PhiSigmaMoment})
we can immediately extract $\phi_{\Sigma}(k)$. Returning to~(\ref{eq:B1-3}) we notice that on the rhs 
again this factor $N_{\Sigma}\left[\phi_{cc}\right]$ appears, and we can write
\begin{eqnarray}
\Phi_{ij}\left(k\right) & = & \left(\omega-N_{n}(i,j)\omega_{n}(k)\right)^{-1}\label{eq:B1-4}\\
 & \times & \left[g^{2}\frac{M_{i}^{2}M_{j}^{2}}{2M_{W}^{2}}N_{\Sigma}\left[\Phi_{cc}\right]+\sqrt{2}\int\frac{d^{2}k'}{(2\pi)^{3}}\,\frac{K_{ij,cc}\left(k,k'\right)\Phi_{cc}\left(k'\right)}{D\left(k',\, M\right)^{2}}\right].\nonumber 
\end{eqnarray}

Finally, substituting~(\ref{eq:B1-4}) into~(\ref{eq:B2}) and combining
with~(\ref{eq:B1-3}),~(\ref{eq:PhiSigma}) we arrive at:
\begin{eqnarray}
 &  & \left(\omega-2\omega_{c}(k)\right)\Phi_{cc}\left(k\right)=\int\frac{d^{2}k'}{(2\pi)^{3}}\,\frac{K_{cc,cc}\left(k,k'\right)\Phi_{cc}\left(k'\right)}{D\left(k',\, M_{W}\right)^{2}}\label{eq:B2-2}\\
 & + & g^{2}\frac{M_{W}^{2}\sqrt{2}}{2}N_{\Sigma}\left[\Phi_{cc}\right]\int\frac{d^{2}k'}{(2\pi)^{3}}\sum_{ij\not=\gamma}\frac{K_{cc,ij}(k,k')(-1)^{N_{3}(i,j)}}{\left(\omega-N_{n}(i,j)\omega_{n}(k')\right)D\left(k',\, M_{i}\right)D\left(k',\, M_{j}\right)}\nonumber \\
 & + & 2\int\frac{d^{2}k'}{(2\pi)^{3}}\int\frac{d^{2}k''}{(2\pi)^{3}}\,\sum_{ij}\frac{K_{cc,ij}(k,k')(-1)^{N_{3}(i,j)}c_{W}^{2N_{Z}(i,j)}s_{W}^{2N_{\gamma}(i,j)}\, K_{ij,cc}\left(k',k''\right)\Phi_{cc}\left(k''\right)}{\left(\omega-N_{n}(i,j)\omega_{n}(k')\right)D\left(k',\, M_{i}\right)D\left(k',\, M_{j}\right)D\left(k'',\, M_{W}\right)^{2}}.\nonumber 
\end{eqnarray}
Equation~(\ref{eq:B2-2}) represents a linear integral equation
for $\Phi_{cc}$. Further simplifications of Eq.~(\ref{eq:B2-2})
are not possible, so we use numerical methods for its analysis.

\begin{figure}
\centering{}\includegraphics[scale=1.2]{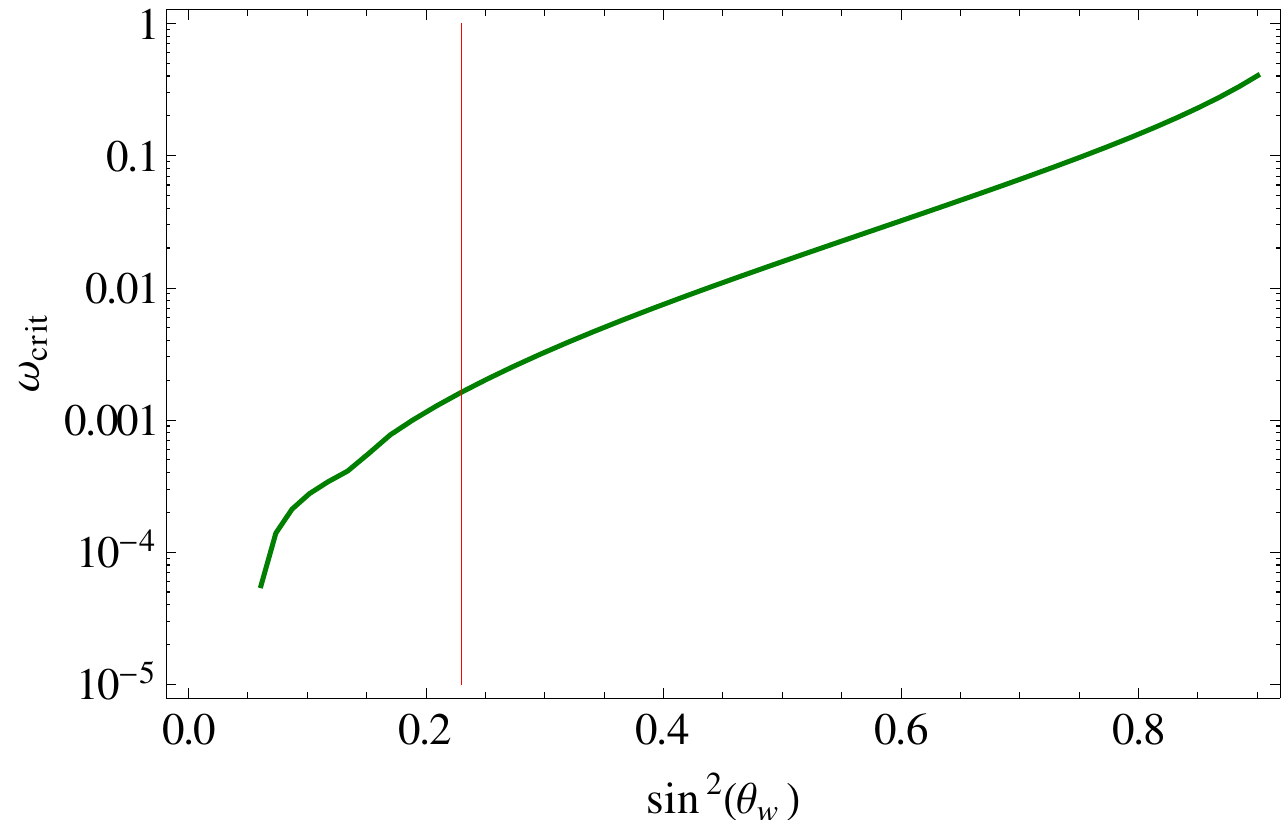} \protect \protect\protect\caption{ \label{fig:PoleBehaviour} Behaviour of the critical eigenvalue $\omega_{{\rm crit}}$
which causes a pole in the second line of Eq.~\ref{eq:PhiSigmaMoment}.
Vertical red line stands for physical value of Weinberg angle $\theta_{W}$. }
\end{figure}

In Section~(\ref{subsec:Generalities}) we have demonstrated that in the case $\theta_{W}=0$,
instead of~(\ref{eq:B2-2}), the isospin zero wave function might
be found as a solution of a much simpler equation~(\ref{eq:Phi_BFKL}).
Albeit~(\ref{eq:B2-2}) does not reduce to (\ref{eq:Phi_BFKL}) in
the limit $\theta_{W}=0$, it is possible to demonstrate that any
solution of~(\ref{eq:Phi_BFKL}) is also a solution of~(\ref{eq:B2-2})
in this limit (see Appendix~(\ref{sec:Equivalence}) for details).
The fact that, in general, the spectrum depends on $\theta_{W}$ can be
seen from the second line in Eq.~(\ref{eq:PhiSigmaMoment}): for
nonzero $\theta_{w}$ due to incomplete cancellation of contributions
of $Z$ and $3$ in denominator appears a term $\sim O(s_{W}^{4})/\omega$.
This contribution for sufficiently small positive $\omega$ leads
to a pole. The position of this pole ($\omega_{0}$) depends on the value
of $\theta_{W}$, as shown in Fig.\ref{fig:PoleBehaviour},
and signals that near the pole there could be a sizeable sensitivity to $\theta_{W}$. A dependence on
$\theta_{w}$ is also contained in the last line of Eq.~(\ref{eq:PhiSigmaMoment}). It turns out that 
net dependence on $\theta_{W}$, for the leading eigenvalue, is negligeable. This result can be traced back to the fact that, in the region of large transverse momenta, the equation aproaches the BFKL
equation. 

\begin{figure}
\centering{}\includegraphics[height=10cm]{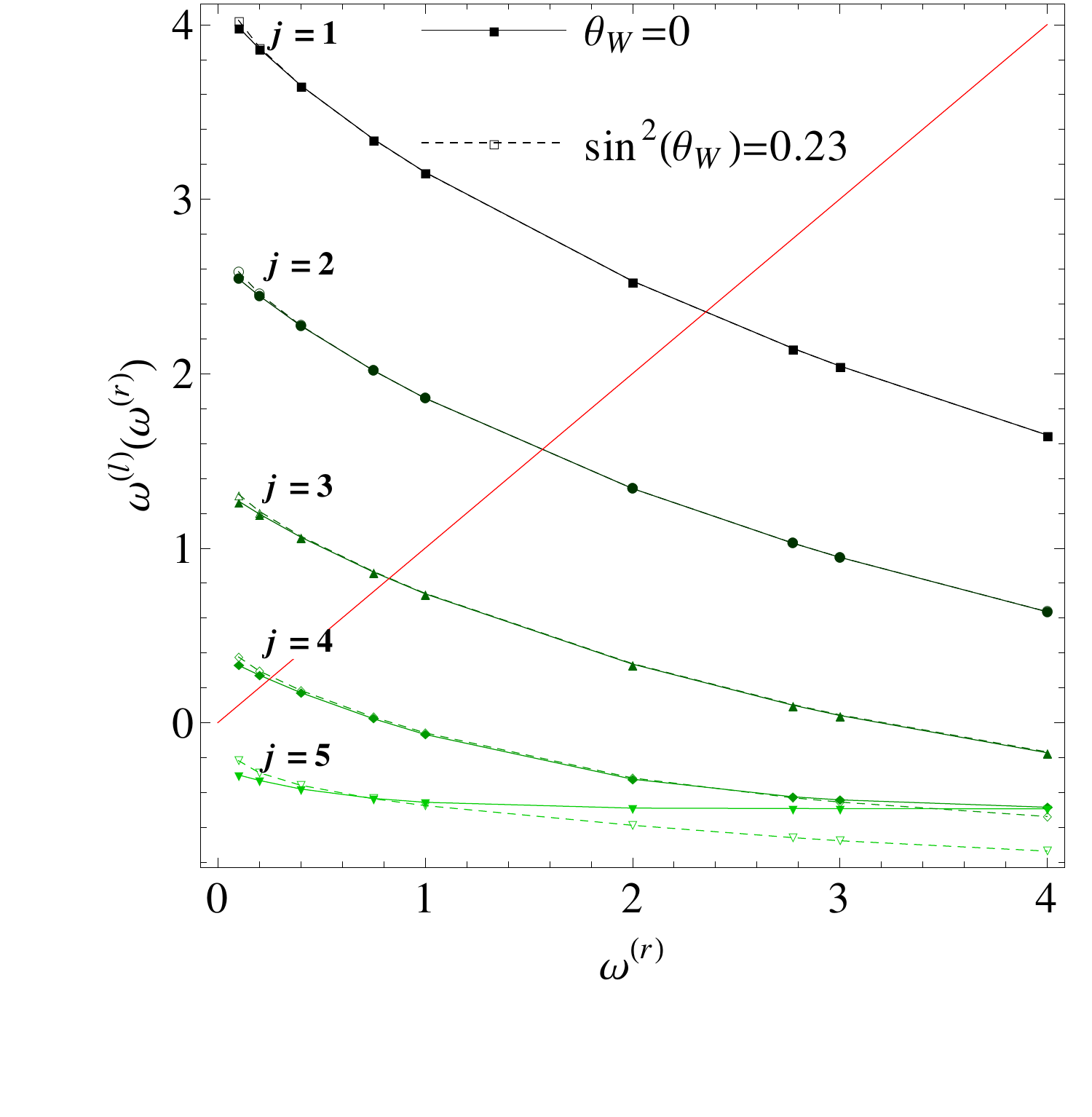} \protect\caption{\label{fig:omegas}Root trajectories $\omega^{(l)}\left(\omega^{(r)}\right)$.
Dashed lines correspond to physical mixing angle $\theta_{W}$. The red
line stands for $\omega^{(l)}=\omega^{(r)}$. We use units $\bar{\alpha}_{{\rm e.w.}}=1$.
For the upper cases j=1,2,3,4 the solid lines ($\theta_W=0$) and the dashed lines (physical mixing angle) are practically indistinguishable, i.e. they are nearly identical.}
\end{figure}

Equation~(\ref{eq:B2-2}) is not a canonical eigenvalue problem. Instead, 
$\omega$ appears in the denominators in the rhs. In order to
solve the equation for $\omega$,  we use the following method: 
\begin{itemize}
\item  On the rhs of~(\ref{eq:B2-2}) we replace $\omega$ everywhere by 
a fixed parameter~%
\footnote{The superscripts $l$ and $r$ stand for ``left'' and ``right''
-hand side substitutions of $\omega$.%
} $\omega^{(r)}$, thus converting Eq.~(\ref{eq:B2-2}) into an ordinary
eigenvalue problem. 
\item We apply the method of Section~(\ref{subsec:NumMethod}) to solve
this eigenvalue problem, and find the corresponding eigenvalues $\omega^{(l)}=\omega^{(l)}\left(\omega^{(r)}\right)$ as functions of the fixed parameter $\omega^{(r)}$.
\item We extract the true eigenvalues of Eq.~(\ref{eq:B2-2}) by solving the algebraic
equation $\omega^{(l)}\left(\omega^{(r)}\right)=\omega^{(r)}$. 
\end{itemize}
Due to the complexity of the equation~(\ref{eq:B2-2}) and to the finite precision
of our numerical evaluation, we cannot extend our lattice up to very
large $\kappa=\frac{k^{2}}{m^{2}}\sim10^{80}$, as we did in Section~(\ref{subsec:NumMethod}).
We fix the minimal and maximal values of $\kappa$ and the number of nodes $N$ as 
\begin{equation}
\kappa_{{\rm min}}=10^{-10},\quad\kappa_{{\rm max}}=10^{15},\quad N=4096.\label{eq:latDef}
\end{equation}
Fig.~\ref{fig:omegas} illustrates that the root trajectories
$\omega^{(l)}\left(\omega^{(r)}\right)$ are homogeneously decreasing
functions of parameter $\omega^{(r)}$, as expected from~(\ref{eq:B2-2}).
For this reason, each trajectory $\omega^{(l)}\left(\omega^{(r)}\right)$
gives rise to only one root $\omega_{j}$. In agreement with our results
of the previous section, for the leading root the effect of a nonzero
mixing angle $\theta_{W}$ is negligible. However, this result is
not universal, and for $\omega\approx0$ the dependence is much stronger.
Finally, in the Table~\ref{tab:EWRoots} we give numerical values
for positive roots.

Due to the relatively small size of the lattice~(\ref{eq:latDef}),
we obtain only four positive roots, in agreement with what was found
in Section~(\ref{subsec:NumMethod}) in Figure~\ref{fig:rootsDep}.
We expect that with an increase of the size of the lattice the intercept
of the leading pole will grow up to its true value $4\,\ln2$, as
well as the distance between the neighboring roots will decrease as
shown in Figure~\ref{fig:rootsDep}. 

\begin{table}
\center %
\begin{tabular}{|c|c|c|}
\hline 
$j$  & $\theta_{W}=0$  & $\sin^{2}\theta_{W}=0.23$\tabularnewline
\hline 
$1$  & 2.335  & 2.331\tabularnewline
\hline 
$2$  & 1.561  & 1.560\tabularnewline
\hline 
$3$  & 0.839  & 0.845\tabularnewline
\hline 
$4$  & 0.245  & 0.262\tabularnewline
\hline 
\end{tabular}\protect\protect\caption{\label{tab:EWRoots}The positive roots $\omega_{j}/\bar{\alpha}_{{\rm e.w.}}$
of~(\ref{eq:B2-2}) evaluated with lattice parameters~(\ref{eq:latDef}). }
\end{table}

Despite of the somewhat low precision of the above estimates , we  clearly see from Table 2 that the first root does not depend on the value of $\theta_W$ for any value of $\theta_W$. As it has been discussed, we know the spectrum at $\theta_W=0$ and we know the first order correction  in $\sin^2\Lb \theta_W\Rb$. Therefore, we can conclude that for $\theta_W\neq 0$ the spectrum of the electro-weak Pomeron is the same as the spectrum of the massless BFKL equation, with the replacement $\bas \to \bar \alpha_{\rm e. w.}$.

\section{Conclusions}

\label{sec:Conclusions} 

In this paper we have analyzed the spectrum of the electroweak BFKL pomeron,
both using perturbative (in $\theta_W$) methods as well as a numerical nonperturbative study 
on the lattice (for $\theta_{W}\not=0$). We found that the leading intercept
important for the high energy behaviour of the amplitudes depends
on $\theta_{W}$ very weakly, and for physical value of $\theta_{W}$
differs from the special case $\theta_{w}=0$ by less than 1\% (see Table 2). However,
for subleading intercepts the dependence on $\theta_{w}$ is more
pronounced. On the other hand, since we have a continuous spectrum, this dependence is not important for the description of the processes.

The leading order intercept is given by $\Delta_{\rm e.w.}\,=\,\omega_{0}\approx\frac{8\alpha_{\mbox{ \tiny e.w}}}{\pi}\ln2\,\approx\, 0.176$,
where $\alpha_{\mbox{ \tiny e.w}}$ is the electroweak fine structure
constant. Numerically, this is a small number. However at, for example,  $W= \sqrt{s} = 30 \,TeV$the contribution of the electroweak Pomeron  $\propto \Lb s/M^2_W\Rb^{\Delta_{\rm e.w.}}$, at $W= \sqrt{s} = 30 \,TeV$ gives already an enhancement of the order of $ \exp\Lb 0.176\,\ln\Lb W^2/M^2_W\Rb\Rb \,\approx\,8$  which is not small. 

\begin{figure}[ht]
\centering 
\begin{tabular}{ccc}
\includegraphics[width=2cm,height=6cm]{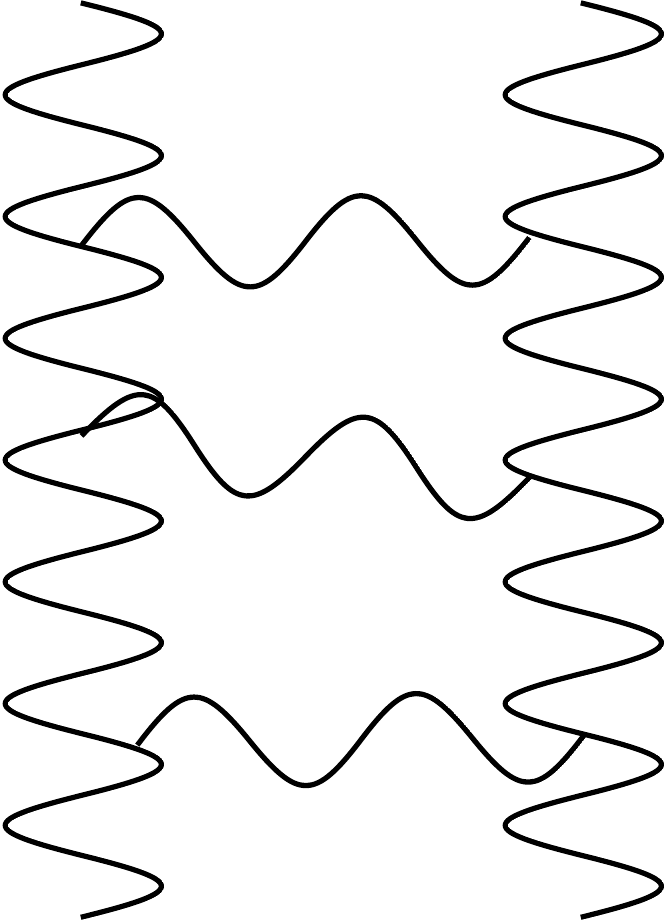}&~~~~~~~~~~~~~~~~~~~~~~~&\includegraphics[width=2cm]{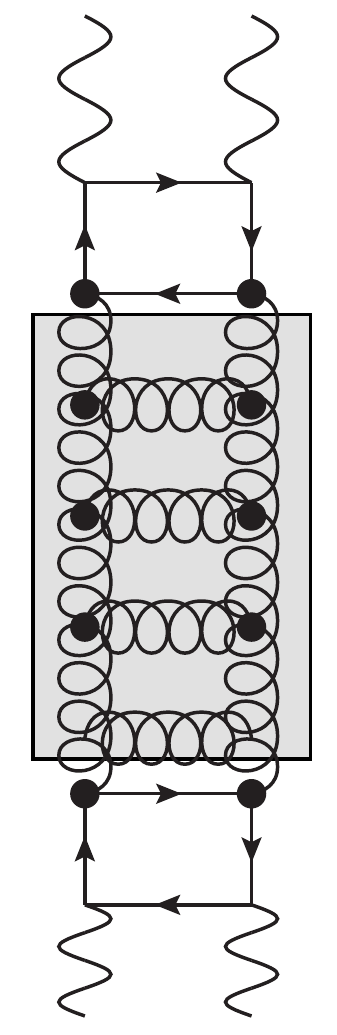}\\
\fig{EWSPom}-a &~~~~~~& \fig{EWSPom}-b\\
\end{tabular}
 \protect\caption{\fig{EWSPom}-a: The contribution of the electroweak Pomeron to an electroweak scattering  process.  \fig{EWSPom}-b: Contribution of the QCD pomeron to the same electroweak processes. External wavy
lines stand for electroweak bosons ($Z,W,\gamma$), the grey blob with
gluon ladder inside stands for the QCD pomeron.}
\label{EWSPom}
\end{figure}

On the other hand, in all practical applications the high energy behavior is dominated by the inclusion of the QCD pomeron, whose intercept is enhanced by a factor $\frac{3}{2}\frac{\alpha_{s}}{\alpha_{\mbox{ \tiny e.w}}}$ (see Figure~\ref{EWSPom}). So the real problem is the coexistence and mixing of the QCD Pomeron with the electroweak one. Let us give an example. First, because of the external coupling the contribution of the QCD Pomeron is suppressed by the factor $\alpha_{\rm e.w.}\,\bas^2$ in comparison to the electroweak Pomeron. Next, for $W = 30\,TeV$  the ratio of the  QCD Pomeron to the electroweak ca be estimated as 
 $\alpha_{\rm e.w.}\,\bas^2\,\exp \Lb \bas \,4\,\ln2\ln\Lb W^2/Q^2\Rb \, -\,0.176\,\ln\Lb W^2/M^2_W\Rb\Rb$ where $Q$ is the virtuality of $W,Z, \gamma$ in  the electroweak process. For $Q= 1 \,GeV$ and $\bas = 0.2$ we see that the ratio is about 20, showing that the QCD Pomerons wins. However, this conclusion maybe a bit premature since the intercept of the QCD Pomeron is strongly affected by the QCD next-to-leading order corrections while such QCD corrections do not change the intercept of the electroweak Pomeron. Pure electroweak corrections are not known. Assuming that the QCD corrections diminish the intercept of the QCD Pomeron by a factor of 2 (as it follows from the phenomenology of DIS), we see that the ratio is about 0.1 showing that we might have a window in the energy in which we are able to measure and to investigate the electroweak Pomeron. The energy $W=30 TeV$ perhaps is too large for the real experiment. At a more realistic energy, $W=3\,TeV$,  the effect of the electroweak Pomeron is smaller than at $W =30\,TeV$. leading to  $ \exp\Lb 0.176\,\ln\Lb W^2/M^2_W\Rb\Rb \,\approx\,3.6$. The contribution of the QCD Pomeron is negligibly small.  Therefore, this energy might give a reasonable compromise between a realistic experimental possibility and seeing an enhancement due to electroweak Pomeron contribution.

We consider our results as an important demonstration that low-energy effects
(like symmetry breaking and mixing of $SU(2)$ with $U(1)$ in physical
bosons) do not affect the leading intercept describing the high energy behavior. A small value of the intercept also ensure that higher order loop corrections remain small and
can be addressed perturbatively. 

An interesting future step of our study is the inclusion of the running
coupling. It is known from the study of the QCD Pomeron~\cite{LIREV}
that the running of $\alpha_s$ leads to a discrete spectrum. The two gluon wave
function belonging to the leading BFKL eigenvalue is concentrated at smaller momenta,
and for this reason we expect a stronger sensitivity to the infrared behaviour
of the theory. In electroweak theory, this should translate into a stronger
sensitivity to a Higgs mechanism and to the value of the mixing angle. 

If the electroweak standard model with one Higgs boson remains the correct theory at LHC energies and beyond (see however, \cite{CMS:2015dxe,ATLAS:2015}) it may be interesting to extrapolate scattering processes and unitarity 
constraints up to energies where strong and electroweak couplings become comparable.
At such energies, QCD and the electroweak sector will mix, and unitarization should affect 
both sectors. Our analysis presented in this paper may provide a starting point for addressing 
such questions.

\section*{Acknowledgements}

We thank our colleagues at UTFSM and at Tel Aviv University for encouraging
discussions. One if us (J.B.) wants to thank the theory group of UTFSM for their generous  hospitality.
This research was  partially supported by the BSF grant   2012124, by    Proyecto Basal FB 0821(Chile) ,  Fondecyt (Chile) grants 1140377 and  1140842,   CONICYT (Chile) grants PIA ACT1406 and ACT1413.
 Powered@NLHPC: This research was partially supported by the supercomputing infrastructure of the NLHPC (ECM-02). Also, we thank Yuri Ivanov for technical support of the USM HPC cluster
where part of evaluations were done.

\appendix

\section{Appendix}

\subsection{Discrete state of the BFKL equation for massive gluon}

\label{sub:discrete}
\begin{figure}[ht]
\centering \includegraphics[width=2cm]{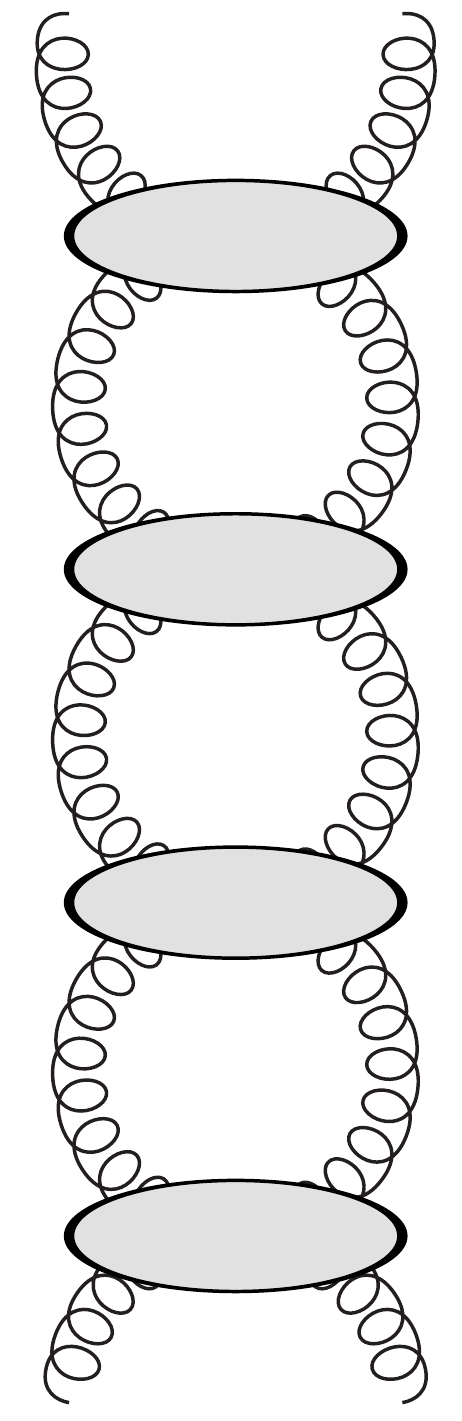} \protect\caption{ The diagrams that lead to the discrete level due to contact term
interaction.}

\label{ct} 
\end{figure}

In~\cite{LLS} we found that in a limit $\theta_{W}\to0$
the spectrum of the BFKL with massive boson is continuous and coincides
with spectrum of massless QCD. Now we would like to demonstrate that
the theory possesses additional discrete levels absent in the massless
limit. However, we overlooked in that paper the existence of a new
discrete level with the intercept $\omega_{discr}\,=\,-\frac{5}{8}\bar{\alpha}_{{\rm e.w.}}$
and with the eigenfunction given by 
\begin{equation}
\phi_{ct}\left(\kappa\right)=\frac{1}{1+\kappa}.\label{eq:phi_ct}
\end{equation}
Plugging the ansatz~(\ref{eq:phi_ct}) into (\ref{NS1}), we may
see that the contributions of the first and the third terms in~(\ref{KKAPPA})
mutually cancel, and the contribution of the second term yields for
the eigenvalue $\omega_{discr}\,=\,-\frac{5}{8}\bar{\alpha}_{{\rm e.w.}}$.
In a diagrammatic language this solution corresponds to a sum of the
diagrams in the Figure~\ref{ct}. 

%%%%%%%%%%%%%%%%%%%%%%%%%%%%%%%%%%%%%%%%% 

However, in a general case $\theta_{W}\not=0$ we can see that solutions
of a form
\begin{align}
\Phi_{ij} & =\frac{c_{ij}}{D(k,M_{i})\, D(k,M_{j})},\\
\Phi_{cc} & =\frac{c}{D(k,M)^{2}}
\end{align}
 do not satisfy the eigenvalue equations~(\ref{eq:B1},\ref{eq:B2}).

\subsection{Numerical results of $\mathbf{\Delta}\mathbf{\omega}_{1}$}

\label{sub:NumericalLattice}In Section~\ref{sub:sw2corr} we argued
that a correction~(\ref{PE1}) is suppressed in the limit $\kappa_{{\rm max}}\to\infty$.
In this section we give for the sake of reference corresponding contributions
of different components in order to support this claim. We use shorthand
notations $\mathbf{\Delta{\boldsymbol{\omega}}_{1}^{\mbox{\ensuremath{\Delta}}K}}$
for the $\mathcal{O}\left(s_{W}^{2}\right)$-corrections to kernels
(\ref{eq:K_ZZcc}-\ref{eq:K_Zncc}, \ref{eq:K_cccc}) and $\mathbf{\Delta{\boldsymbol{\omega}}_{1}^{\mbox{\ensuremath{\Delta}}P}}$
due to $\mathcal{O}\left(s_{W}^{2}\right)$-corrections to propagators.
Using an upper cutoff as in Section~\ref{subsec:NumMethod}, we found
{\footnotesize{}
\begin{equation}
\mathbf{\Delta{\boldsymbol{\omega}}_{1}^{\mbox{\ensuremath{\Delta}}K}}=s_{W}^{2}\,10^{-4}\times\left(\begin{array}{cccccccc}
 & ZZ & \tilde{n}\tilde{n} & \{Z\tilde{n}\} & 3\,3 & \{Z3\} & \{\tilde{n}3\} & cc\\
ZZ~ & 1.10 & 0.55 & 1.65 & 0.55 & -1.65 & -1.65 & 0.48\\
\tilde{n}\tilde{n} & 0.55 & 0. & 0.55 & 0. & -0.55 & -0.55 & 0.\\
\{Z\tilde{n}\} & 0.82 & 0.27 & 1.10 & 0.27 & -1.10 & -1.10 & 0.24\\
3\,3 & 0.55 & 0. & 0.55 & 0. & -0.55 & -0.55 & 0.24\\
\{Z3\} & -0.82 & -0.27 & -1.10 & -0.27 & 1.10 & 1.10 & -0.24\\
\{\tilde{n}3\} & -0.82 & -0.27 & -1.10 & -0.27 & 1.10 & 1.10 & -0.24\\
cc & 0.48 & 0. & 0.48 & 0. & -0.48 & -0.48 & -24
\end{array}\right)\label{DOMDC}
\end{equation}
}and

{\footnotesize{}
\begin{equation}
\mathbf{\Delta{\boldsymbol{\omega}}_{1}^{\mbox{\ensuremath{\Delta}}P}}=s_{W}^{2}\,10^{-4}\times\left(\begin{array}{cccccccc}
 & ZZ & \tilde{n}\tilde{n} & \{Z\tilde{n}\} & 3\,3 & \{Z3\} & \{\tilde{n}3\} & cc\\
ZZ~ & -3.16 & 0. & -2.58 & 0. & 2.58 & 0. & 0.\\
\tilde{n}\tilde{n} & 0.82 & 0.27 & 1.10 & 0.27 & -1.10 & -1.10 & 0.24\\
\{Z\tilde{n}\} & 0.82 & 0.27 & 1.10 & 0.27 & -1.10 & -1.10 & 0.24\\
3\,3 & 0.82 & 0.27 & 1.10 & 0.27 & -1.10 & -1.10 & 0.24\\
\{Z3\} & 0.82 & 0.27 & 1.10 & 0.27 & -1.10 & -1.10 & 0.24\\
\{\tilde{n}3\} & 0.82 & 0.27 & 1.10 & 0.27 & -1.10 & -1.10 & 0.24\\
cc & L-8.90 & 0. & 0.89-L & 0. & L-0.89 & 0. & 0.
\end{array}\right)\label{DOMPC}
\end{equation}
} where $L=-0.53\times10^{4}$. We can see that, indeed, all corrections
except those~$\sim L$ are small and vanish in the limit $\kappa_{{\rm max}}\to\infty$.
The components $\sim L$ vanish in a full sum, which finishes the
proof that the corrections are small.

\subsection{Proof that (\ref{eq:B2-2}) includes (\ref{eq:Phi_BFKL}) for $\theta_{W}=0$ }

\label{sec:Equivalence} In this section we demonstrate explicitly
that for the case $\theta_{W}=0$ any solution of Equation~(\ref{eq:Phi_BFKL})
does satisfy the equation (\ref{eq:B2-2}). Indeed, we may rewrite~(\ref{eq:Phi_BFKL})
as
\begin{equation}
\int d^{2}k'\, K_{em}\left(k,k'\right)\frac{\Phi_{cc}\left(k'\right)}{D^{2}\left(k'\right)}=-g^{2}\frac{5M_{W}^{2}}{8}\int d^{2}k'\,\frac{\Phi_{cc}\left(k'\right)}{D^{2}\left(k'\right)}+\frac{\left(\omega-2\omega(k)\right)}{4}\Phi_{cc}\left(k\right).
\label{eq:KPhi}
\end{equation}
Since in the limit $\theta_{W}=0$ the kernels $K{}_{ij,cc}$ and
$K_{cc,cc}$, up to a constant, are proportional to $K_{em}\left(k,k'\right)$,
\begin{eqnarray}
 &  & K_{cc,cc}\left(k,k'\right)=-g^{2}M_{W}^{2}+2K_{em}\left(k,k'\right)\\
 &  & K_{cc,nn}(k,k')=-\frac{3g^{2}M_{W}^{2}}{2}+2K_{em}\left(k,k'\right),
\end{eqnarray}
we can use~(\ref{eq:KPhi}) to evaluate explicitly the convolutions.
Eq. (\ref{eq:B2-2}) in the limit $\theta_{W}=0$ is given by
\begin{eqnarray}
\left(\omega-2\omega_{n}(k)\right)D\left(k,\, M_{W}\right)^{2}\phi_{cc}\left(k\right) & = & \int d^{2}k'\,\left[-M_{W}^{2}+2K_{em}\left(k,k'\right)\right]\frac{\Phi_{cc}\left(k'\right)}{D^{2}\left(k'\right)}\label{eq:B2-3}\\
 & + & \frac{M_{W}^{2}}{2}\frac{f(k)\int d^{2}k'\, f(k')\Phi_{cc}\left(k'\right)/D^{2}\left(k'\right)}{1-\frac{M_{W}^{2}}{2}\int\frac{d^{2}k}{\left(\omega-2\omega_{n}(k)\right)D^{2}\left(k\right)}}\nonumber \\
 & + & 2\int d^{2}k'\int d^{2}k''\,\frac{K_{cc,nn}(k,k')\, K_{nn,cc}\left(k',k''\right)}{\left(\omega-2\omega_{n}(k')\right)D^{2}\left(k'\right)}\frac{\Phi_{cc}\left(k''\right)}{D^{2}\left(k''\right)},\nonumber 
\end{eqnarray}
where 
\begin{equation}
f(k)=\sqrt{2}\int d^{2}k'\frac{K_{cc,nn}(k,k')}{\left(\omega-2\omega_{n}(k')\right)D^{2}\left(k'\right)}.
\end{equation}
A straightforward application of~(\ref{eq:KPhi}) after some algebra
yields 
\begin{eqnarray}
\int d^{2}k'\, f(k')\,\frac{\Phi_{cc}\left(k'\right)}{D^{2}\left(k'\right)} & = & \frac{\sqrt{2}}{2}\int d^{2}k'\,\frac{\Phi_{cc}\left(k'\right)}{D^{2}\left(k'\right)}\left[1-\frac{M_{W}^{2}}{2}\int\frac{d^{2}k}{\left(\omega-2\omega_{n}(k)\right)D^{2}\left(k\right)}\right],\label{eq:E1}
\end{eqnarray}
\begin{eqnarray}
\frac{M_{W}^{2}}{2}\frac{f(k)\int d^{2}k'\, f(k')\Phi_{cc}\left(k'\right)/D^{2}\left(k'\right)}{1-\frac{M_{W}^{2}}{2}\int\frac{d^{2}k}{\left(\omega-2\omega_{n}(k)\right)D\left(k\right)D\left(k\right)}} & = & \frac{M_{W}^{2}}{2}\int d^{2}k''\,\frac{\Phi_{cc}\left(k''\right)}{D^{2}\left(k''\right)}\int d^{2}k'\frac{\left[-\frac{3M_{W}^{2}}{2}+2K_{em}\left(k,k'\right)\right]}{\left(\omega-2\omega_{n}(k')\right)D^{2}\left(k'\right)},\label{eq:E2}
\end{eqnarray}
\begin{eqnarray}
2\int d^{2}k'\int d^{2}k''\,\frac{K_{cc,nn}(k,k')\, K_{nn,cc}\left(k',k''\right)}{\left(\omega-2\omega_{n}(k')\right)D\left(k'\right)D\left(k'\right)}\,\frac{\Phi_{cc}\left(k''\right)}{D^{2}\left(k''\right)} & = & -g^{2}\frac{M_{W}^{2}}{4}\int d^{2}k'\,\frac{\Phi_{cc}\left(k'\right)}{D^{2}\left(k'\right)}\nonumber \\
 & + & \frac{\left(\omega-2\omega_{n}(k)\right)}{2}\Phi_{cc}\left(k\right)\label{eq:E3}\\
 & - & g^{2}\frac{M_{W}^{2}}{2}\int d^{2}k'\,\frac{\left[-\frac{3M_{W}^{2}}{2}+2K_{em}\left(k,k'\right)\right]}{\left(\omega-2\omega_{n}(k')\right)D^{2}\left(k'\right)}\nonumber \\
 & \times & \int d^{2}k''\frac{\Phi_{cc}\left(k''\right)}{D^{2}\left(k''\right)},\nonumber 
\end{eqnarray}
and 
\begin{eqnarray}
\int d^{2}k'\, K_{cc,cc}\left(k,k'\right)\frac{\Phi_{cc}\left(k'\right)}{D^{2}\left(k'\right)} & = & \left[\frac{\left(\omega-2\omega_{n}(k)\right)\Phi_{cc}\left(k\right)}{2}+g^{2}\frac{M_{W}^{2}}{4}\int d^{2}k''\frac{\Phi_{cc}\left(k''\right)}{D^{2}\left(k''\right)}\right].
\label{eq:E4}
\end{eqnarray}
After summation of~(\ref{eq:E1}-\ref{eq:E4}) we recover the
lhs of~(\ref{eq:B2-3}).

%%%%%%%%%%%%%%%%%%%%%%%%%%%%%%%%%%%%%%%%%%%%%%%%% 

\subsection{Large transverse momenta}
\label{subsec:large}
%%%%%%%%%%%%%%%%%%%%%%%%%%%%%%%%%%%%%%%%%%%%%%%%%%%%%%%%%%%%%%%
In this appendix we show how \eq{eq:B2-2}  has the same solution as  the BFKL equation at large values of $\kappa\, >\,M^2_Z$, which we have discussed in section 3 (see \eq{BFKL} and \eq{BFKLLIM}).  For large $\kappa$ we can re-write  functions $\Phi_{i j}$ and $\Phi_{cc}$ of section 4  in the form
\ba \label{C1}
\Phi_{ij}\Lb \kappa\Rb\,&\xrightarrow{\kappa\,\gg\,M^2_Z}&\,D\Lb \kappa, M_i\Rb\,\phi_W\Lb \kappa\Rb\,\,=\,\,\kappa\, \kappa^{-\h + i \nu}\,,\nn\\
\Phi_{cc}\Lb \kappa\Rb\,&\xrightarrow{\kappa\,\gg\,M^2_W}&\,D\Lb \kappa, M_W\Rb\,\phi_W\Lb \kappa\Rb\,\,=\,\,\frac{1}{\kappa}\, \kappa^{-\h + i \nu}\, ,
\ea
where $\phi_W\Lb \kappa\Rb$ is the eigenfunction of the BFKL equation (see \eq{BFKL}).
In \eq{C1} we first calculate the last term of \eq{eq:B2-2}
  \be \label{C11}
  2\int\frac{d^{2}k'}{(2\pi)^{3}}\int\frac{d^{2}k''}{(2\pi)^{3}}\,\sum_{ij}\frac{K_{cc,ij}(k,k')(-1)^{N_{3}(i,j)}c_{W}^{2N_{Z}(i,j)}s_{W}^{2N_{\gamma}(i,j)}\, K_{ij,cc}\left(k',k''\right) \Phi_{cc}\left(k''\right)}{\left(\omega-N_{n}(i,j)\omega_{n}(k')\right)\,D\left(k',\, M_{i}\right) \,D\left(k',\, M_{j}\right)\,D^2\Lb k'',M_W\Rb}.
\ee
 First, in the sum over $i, j$ we consider the terms with $i = j = n$, which after substituting \eq{C1} takes the form
 \ba \label{C2}
 &&  2\int\frac{d^{2}k'}{(2\pi)^{3}}\int\frac{d^{2}k''}{(2\pi)^{3}} \frac{K_{cc,nn}\Lb k, k'\Rb\,K_{nn, cc}\Lb k', k''\Rb\,\phi_{W}}{\Lb \omega\,-\,2 \omega_n\Lb k' \Rb\Rb \,D\Lb k', M_W\Rb}\,=\\
 &&  2\int\frac{d^{2}k'}{(2\pi)^{3}}\int\frac{d^{2}k''}{(2\pi)^{3}} \frac{K_{cc,nn}\Lb k, k'\Rb\Lb \h \Lb \omega\,-\,2\,\omega_n\Lb k'\Rb\Rb\,\phi_W\Lb k'\Rb\,- \frac{3}{2}\phi_W\Lb k''\Rb/k''^2\Rb}{\Lb \omega\,-\,2\,\omega_n\Lb k'\Rb\Rb D^2\Lb k',M_W\Rb} \nn\\
 &\xrightarrow{k,k',k'' \gg M_W}& \int\frac{d^{2}k'}{(2\pi)^{3}}\,\frac{K_{cc,nn}\Lb k, k'\Rb }{D\Lb k',M_W\Rb}\phi_W\Lb k'\Rb\,\xrightarrow{k,k'' \gg M_W}\, 2  \int\frac{d^{2}k'}{(2\pi)^{3}}\,\frac{K_{em}\Lb k, k'\Rb}{ D\Lb k',M_W\Rb}\phi_W\Lb k'\Rb .\nn
  \ea
   In the second line of \eq{C2} we use the explicit form of the kernel $K_{nn, cc}\Lb k', k''\Rb$ (see \eq{kijcc} ) while  in the last line we
   neglect the terms that are of the order of $1/k^2$ at large $k$ in comparison $\phi_W\Lb k \Rb$. Note, that this term is the same as the first term in \eq{eq:B2-2}.
   Replacing $K_{cc,i j}\Lb k,k'\Rb$ and $K_{ij,cc}\Lb k',k''\Rb$ by $2\, K_{em}\Lb k,k'\Rb$ and $2 \,   K_{em}\Lb k',k''\Rb $, respectively, at large $k, k'$ and   $k''$ we can see that the sum over $n, j$  with $j=Z,3,\gamma$
   can be reduced to
   \be \label{C3}
 8\int\frac{d^{2}k'}{(2\pi)^{3}}\int\frac{d^{2}k''}{(2\pi)^{3}} \frac{K_{em}\Lb k, k'\Rb\,K_{em}\Lb k', k''\Rb\,\phi_{cc}\Lb k''\Rb}{\Lb \omega\,-\,\omega_n\Lb k' \Rb\Rb \,k'^4}\,\underbrace{\Big\{ c^2_W - 1 + s^2_W\Big\}}_{ = 0}.
 \ee
      
Summation over $i,j$  $i = Z, 3,\gamma $ and $j = Z, 3, \gamma$ reduces to the replacement $\omega\,-\,\omega_n\Lb k' \Rb    \,\to\,\omega $ and $\{\dots\} \,\to\,\{ c^4_W + 1 + s^4_W + 2 c^2_W\,s^2_W   - 2 c^2_W - 2 s^2_W \} = 0$, in \eq{C3}.  Therefore, the only term which contributes at large momenta is the term with$i j$= $n n$, and  \eq{eq:B2-2} reduces to the BFKL equation (see \eq{BFKL}) since the second term in \eq{eq:B2-2} vanishes in this kinematic region.  It is instructive to note that we have proven the equivalence of \eq{eq:B2-2} and \eq{BFKL} without assuming that $| \kappa - \kappa'|\,\gg\,M^2_W$.

%%%%%%%%%%%%%%%%%%%%%%%%%%%%%%%%%%%%%%%%%%%%%%%%% 

\end{document}